%% file: draft.tex
\documentclass[prd,aps,a4paper,superscriptaddress,twocolumn,nofootinbib]{revtex4}
\usepackage{graphicx}
\usepackage{color}
\usepackage{dcolumn}
\usepackage{bm}
\usepackage{slashed}
\usepackage{amsmath}
\usepackage{latexsym}
\usepackage{amssymb}
\usepackage{mathrsfs}
\usepackage{amsfonts}
\usepackage{url}
\usepackage{longtable}
\allowdisplaybreaks
\begin{document}
\title{Accuracy of numerical relativity waveforms with respect to space-based gravitational wave detectors}

\author{Zun Wang} 
\affiliation{Institute for Frontiers in Astronomy and Astrophysics, Beijing Normal University, Beijing 102206, China}
\affiliation{Department of Astronomy, Beijing Normal University, Beijing 100875, China}
\author{Junjie Zhao} 
\affiliation{Institute for Frontiers in Astronomy and Astrophysics, Beijing Normal University, Beijing 102206, China}
\affiliation{Department of Astronomy, Beijing Normal University, Beijing 100875, China}
\author{Zhoujian Cao
\footnote{corresponding author}} \email[Zhoujian Cao: ]{zjcao@amt.ac.cn}
\affiliation{Institute for Frontiers in Astronomy and Astrophysics, Beijing Normal University, Beijing 102206, China}
\affiliation{Department of Astronomy, Beijing Normal University, Beijing 100875, China}
\affiliation{School of Fundamental Physics and Mathematical Sciences, Hangzhou Institute for Advanced Study, UCAS, Hangzhou 310024, China}

\begin{abstract}
The same to laser interferometer gravitational-wave observatory (LIGO), matched filtering technique will be critical to data analysis of gravitational wave detection by space-based detectors including LISA, Taiji and Tianqin. Waveform templates are basis for such matched filtering technique. In order to construct ready-to-use waveform templates, numerical relativity waveforms are start point. So the accuracy issue of numerical relativity waveforms is critically important. There are many investigations about this issue with respect to LIGO. But unfortunately there are few results on this issue with respect to space-based detectors. The current paper investigates this problem. Our results indicate that the existing numerical relativity waveforms are as accurate as 99\% with respect to space-based detectors including LISA, Taiji and Tianqin. Such accuracy level is comparable to the one with respect to LIGO.
\end{abstract}

\maketitle

\section{Introduction}
Since the first successful detection of gravitational waves in 2015, about 100 gravitational wave events have been reported. All these events are found by matched filtering technique. Till now matched filtering technique is still the standard data analysis trick for gravitational waves detection \cite{2021arXiv211103606T,2023arXiv230203676T}. As expected, the situation will be also true for space-based detectors in the near future \cite{Miller2019,2023LRR....26....2A}.

In order to let the matched filtering technique work, accurate and complete waveform templates are needed \cite{2007arXiv0711.1115J,2021hgwa.bookE..43K,2022NatAs...6.1356S,RevModPhys.94.025001}.
Besides the above mentioned matched filtering method, there is time-frequency excess power identification method \cite{2017arXiv171009256C}. But if accurate waveform is available and aid the analysis, the method will become more efficient \cite{PhysRevD.98.024028}. In addition, machine learning method is a new trick to treat gravitational wave data \cite{2020arXiv200503745C,2023PhRvD.107b3021S,PhysRevD.101.104003,PhysRevD.103.024040,PhysRevD.105.083013,PhysRevD.107.063029}. A data set consisting a large amount of gravitational wave samples is critical important to let the machine learning method work well. Since the real gravitational wave events are too few to play the role of such a data set, accurate waveform templates are needed. In a short summary, no matter what kind of data analysis means are taken, waveform templates are very important to gravitational wave data analysis.

A waveform template means the accurate waveform with respect to time or frequency when a set of source parameters are specified. That is to say a waveform template is valid only for a class of source falling in a specified parameters range. Till now gravitational wave astronomy community only understand binary compact object systems well. Consequently only waveform template of binary systems is available currently. This fact also explains to some extent why only events of coalescence of binary objects are observed till now. In contrast, current detection ability for supernovae gravitational waves is quite weak \cite{PhysRevD.101.084002}. Roughly the detection horizon is just 1kpc (Fig.~5(a) of \cite{PhysRevD.101.084002}). The partial reason for such fact is the lack of accurate waveform template for supernovae gravitational waves.

Before the breakthrough of numerical relativity \cite{PhysRevLett.95.121101,PhysRevLett.96.111101,PhysRevLett.96.111102,PhysRevD.78.124011,Zhao2020}, the waveform template problem is treated mainly through post-Newtonian approximation \cite{PhysRevD.49.2658}. As an enhanced post-Newtonian approximation method, effective one body theory shows better convergent behavior \cite{PhysRevD.59.084006}. After the success of numerical relativity simulation of binary black hole merger, the complete inspiral-merger-ringdown behavior is revealed. The power of effective one body theory to describe the waveform of coalescence of binary object is verified \cite{PhysRevD.76.104049}. After that the numerical relativity waveforms are extensively used to construct waveform templates for coalescence of binary objects.

Till now, there are a bundle of waveform templates for coalescence of binary objects available in the LIGO data analysis software. Among kinds of waveform template models including EOBNR series \cite{PhysRevD.95.044028_SEOBNRv4,PhysRevD.101.101501_TEOBeccc,PhysRevD.96.044028,PhysRevD.101.044049,2022CQGra..39c5009L,IJMPD.32.2350015}, IMRphenom series \cite{2020PhRvD.102f4001P,PhysRevD.102.064002}, numerical relativity surrogate models \cite{PhysRevD.96.024058,PhysRevResearch.1.033015,PhysRevD.103.064022} and others, the numerical relativity waveforms are bases for the waveform templates construction. When people talk about the accuracy of a waveform model, the numerical relativity waveforms are treated as the standard answer. So the accuracy of numerical relativity waveforms themselves are critically important to waveform template construction.

Both ground-based and space-based gravitational wave detectors utilize matched filtering techniques for detecting gravitational waves. Therefore, when calculating the accuracy of the templates, the formulas used are very similar to those of the matched filtering technique. Considering the detector noise is crucial when using the matched filtering technique to search for signals. Hence, when calculating the accuracy of the waveform templates, it is necessary to take into account the sensitivity of different detectors. The noise characteristics of space-based gravitational wave detectors differ significantly from those of ground-based detectors. Therefore, the purpose of this article is to investigate whether the numerical relativity waveform's accuracy can meet the requirement of space-based detectors.
When ones discuss the accuracy of a waveform model, a specific detector should be referred. The accuracy issue of numerical relativity waveforms has been extensively studied against advanced LIGO detectors \cite{2019CQGra..36s5006B}. But this issue has not been investigated against space-based detectors. The current paper aims to do such an investigation and lays down a foundation of waveform template construction for space-based detectors.

In the next section we introduce the waveform accuracy estimation method. After that we apply the method in Sec.~\ref{secIII} to calculate waveform accuracy of the waveforms of SXS numerical relativity catalog. LISA, Taiji and Tianqin detectors are all considered. Finally the summary and conclusion are given in the last section.
\section{Matching factor and accuracy indicator}
Following the idea of matched filtering data analysis trick, matching factor has been extensively used to quantify how close between two given waveforms. With respect to a detector sensitivity $S(f)$ which describes the one sided power spectrum of the detector noise, the matching factor of two real waveforms $h_1(t)$ and $h_2(t)$ can be expressed as
\begin{align}
{\rm FF}&\equiv\max_{t}\frac{\langle h_1|h_2\rangle}{\|h_1\|\cdot\|h_2\|},\label{equation1}\\
\langle h_1|h_2\rangle&=2\int_{f_{\rm low}}^{f_{\rm up}}\frac{\tilde{h}_1\tilde{h}_2^*+\tilde{h}_1^*\tilde{h}_2}{S(f)}df,\label{equation2}\\
\|h\|&\equiv\sqrt{\langle h|h\rangle},
\end{align}
where the ``$\tilde{(\cdot)}$" means the Fourier transformation, the ``${}^*$" means taking the complex conjugate, and the maximum is taken with respect to the time shift to align the two waveforms. $(f_{\rm low},f_{\rm up})$ corresponds to the frequency band where the two waveforms should be compared. Within \text{PyCBC} software \cite{PyCBC}, the command line `\text{pycbc.filter.matchedfilter.match}' can be used to do the above calculation of the matching factor ${\rm FF}$. Together with the value of the matching factor, the time shift is also returned by the command line.

For a theoretical waveform template, two polarization waveforms will be given $h_+(t)$ and $h_\times(t)$. Usually people are used to the complex waveform defined as
\begin{align}
h\equiv h_+-ih_\times.
\end{align}
Then similar matching factor to (\ref{equation1}) can be defined to quantify the closeness between two complex waveforms. The only difference to (\ref{equation1}) is the maximum should be taken with respect to the initial phase besides the shifted time. The initial phase describes the phase difference between the two polarization modes $h_+(t)$ and $h_\times(t)$ at the initial time.

Assume we have two complex waveforms $h_{1,2}=h_{1,2+}-ih_{1,2\times}$, the linearity of inner product (\ref{equation2}) results in
\begin{align}
\langle h_1|h_2\rangle&=\langle h_{1+}|h_{2+}\rangle+\langle h_{1\times}|h_{2\times}\rangle\nonumber\\
&-i\langle h_{1+}|h_{2\times}\rangle-i\langle h_{1\times}|h_{2+}\rangle.
\end{align}
In the mean time Eq.~(\ref{equation2}) can be equivalently expressed as
\begin{align}
\langle h_1|h_2\rangle&=4\Re\int\frac{\tilde{h}_1\tilde{h}_2^*}{S(f)}df,
\end{align}
where $\Re$ means taking the real part. So we have
\begin{align}
\langle h_1|h_2\rangle&=\langle h_{1+}|h_{2+}\rangle+\langle h_{1\times}|h_{2\times}\rangle,
\end{align}
which corresponds to
\begin{align}
\max_{t,\phi}\langle h_1|h_2\rangle&=\max_{t_+}\langle h_{1+}|h_{2+}\rangle+\max_{t_\times}\langle h_{1\times}|h_{2\times}\rangle.
\end{align}
Here $t$ means the time shift for the complex waveform, $\phi$ means the initial phase difference of the two polarization modes, and $t_{+,\times}$ are the time shifts for the two polarization waveforms. Due to the above relations, we have
\begin{align}
&{\rm FF}=\frac{{\rm FF}_+\|h_{1+}\|\cdot\|h_{2+}\|+{\rm FF}_\times\|h_{1\times}\|\cdot\|h_{2\times}\|}{\|h_1\|\cdot\|h_2\|},\\
&{\rm FF}_+\equiv\max_{t}\frac{\langle h_{1+}|h_{2+}\rangle}{\|h_{1+}\|\cdot\|h_{2+}\|},\\
&{\rm FF}_\times\equiv\max_{t}\frac{\langle h_{1\times}|h_{2\times}\rangle}{\|h_{1\times}\|\cdot\|h_{2\times}\|}.
\end{align}
That is to say we can calculate the matching factors ${\rm FF}_+$ and ${\rm FF}_\times$ for two polarization modes individually, then use the above equation to combine the final matching factor we wanted. In the current work we follow this way and use PyCBC tool to calculate the matching factor.

Similar to any other computing science topics, the only errors involved in numerical relativity include truncation errors and round off errors. Truncation errors are due to the numerical approximation of derivatives. Round off errors are due to the memory limit of computers. In practice, ones need to make sure the real calculation dominated by truncation errors. Consequently the final error related to the numerical solution is proportional to some power of the resolution used in the numerical calculation. The power index is nothing but the convergence order of the involved numerical algorithm. So we can use the difference between the results of two different resolutions to quantitatively estimate the error of the numerical solution.

In the current work we use the matching factor between the two numerical relativity waveforms of two different resolutions to quantify the accuracy of the numerical relativity waveforms. Specifically to the SXS waveform catalogs \cite{SXSBBH,2019CQGra..36s5006B}, the finest and second finest resolutions are used.

\section{Accuracy of numerical relativity waveforms}\label{secIII}

\subsection{Fourier transforms of numerical relativity waveforms}
Numerical relativity (NR) waveforms are presented in time domain. In order to calculate the matching factor explained in the last section, we need transform these waveforms to frequency domain. In practice, we use fast Fourier transformation to get the waveforms in frequency domain.

In order to reduce the Gibbs effect and spectral leakage resulting from truncation in the time domain, we apply the Plank window $\sigma_{T}(t)$ to the time domain waveform before the Fourier transformation. The Plank window $\sigma_{T}(t)$ is set as \cite{chu2016accuracy,mckechan2010tapering}
\begin{equation}
    \sigma(t)=\left\{\begin{array}{lr}
    0, & t<t_{1} \\
    \sigma_{\text {start }}(t), & t_{1} \leq t<t_{2} \\
    1, & t_{2} \leq t<t_{3} \\
    \sigma_{\text {end }}(t), & t_{3} \leq t<t_{4} \\
    0, & t_{4} \leq t
    \end{array}\right.
    \label{eq:plankwin}
\end{equation}
where $\sigma_{start}$ is the segment that smoothly increases from 0 to 1 between $t_1$ and $t_2$ , and $\sigma_{end}$  is the segment that smoothly decreases from 1 to 0 between $t_3$ and $t_4$ :
\begin{equation}
    \begin{array}{l}
    \sigma_{\text {start }}(t)=\left[\exp \left(\frac{t_{2}-t_{1}}{t-t_{1}}+\frac{t_{2}-t_{1}}{t-t_{2}}\right)+1\right]^{-1}, \\
    \sigma_{\text {end }}(t)=\left[\exp \left(\frac{t_{3}-t_{4}}{t-t_{3}}+\frac{t_{3}-t_{4}}{t-t_{4}}\right)+1\right]^{-1} .
    \end{array}
    \label{eq:sigma}
\end{equation}
Then, We further zero pad the waveform to the nearest power of 2.

\subsection{Frequency range of numerical relativity waveforms}

\begin{figure}
\begin{tabular}{c}
\includegraphics[width=0.45\textwidth]{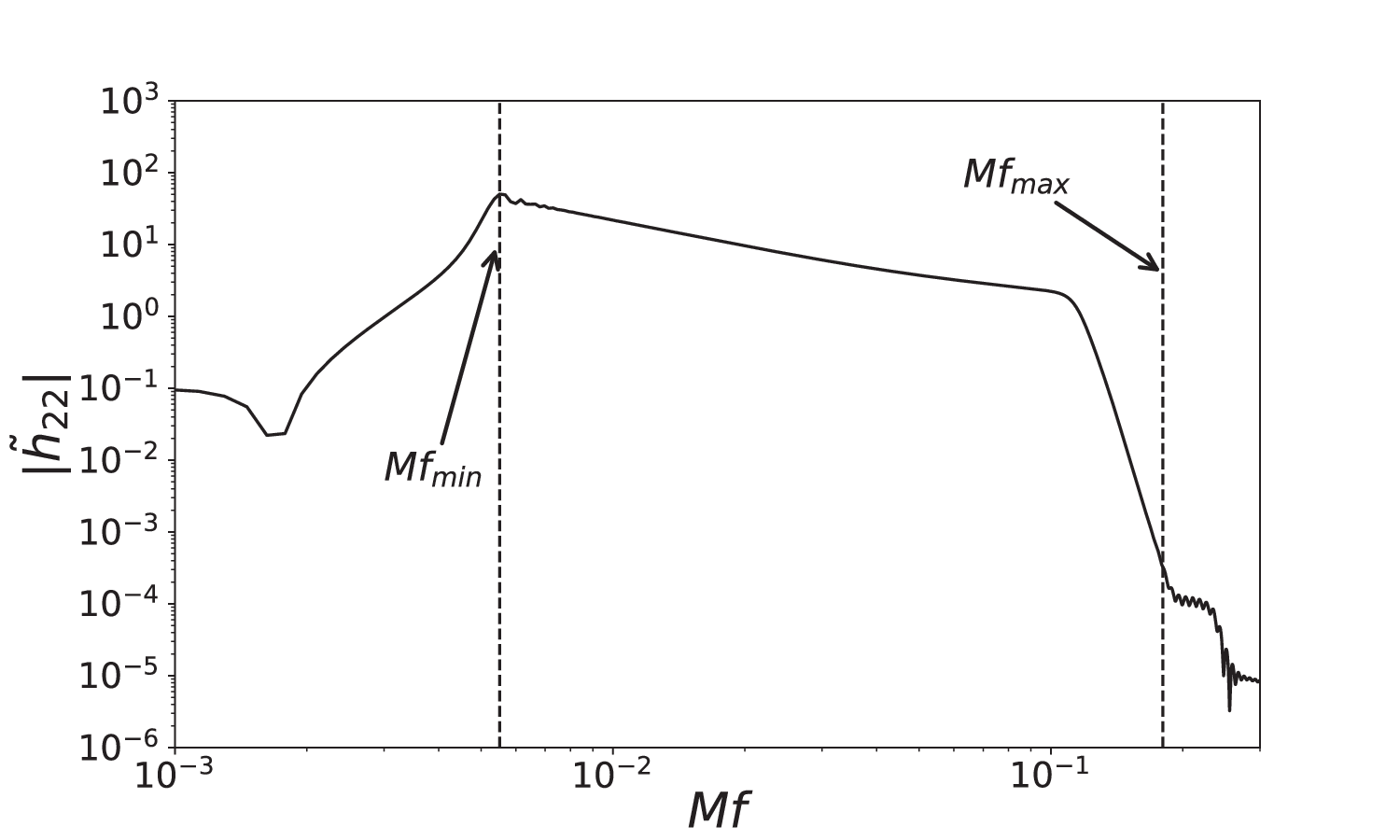}
\end{tabular}
\caption{Frequency waveform of SXS:BBH:2106. This waveform corresponds to a quasi-circular coalescing binary black hole system with mass ratio 1, dimensionless spin $\vec{\chi}_1=(0,0,0.8998)$ and $\vec{\chi}_2=(0,0,0.5)$. In the plot, $M$ means the total mass of the binary. The horizontal axis has no special meaning. It just indicates different NR simulations.}\label{fig1}
\end{figure}

Typically we get waveforms in frequency domain like the one shown in Fig.~\ref{fig1}. Apparently only the part between the two vertical dash lines is reliable. The left vertical line corresponds to the lowest frequency $f_{\rm min}$ of the numerical relativity waveform which is determined by the length of the waveform. The right vertical line corresponds the highest frequency $f_{\rm max}$ where the numerical error begins to dominate.

\begin{figure}
\begin{tabular}{c}
\includegraphics[width=0.45\textwidth]{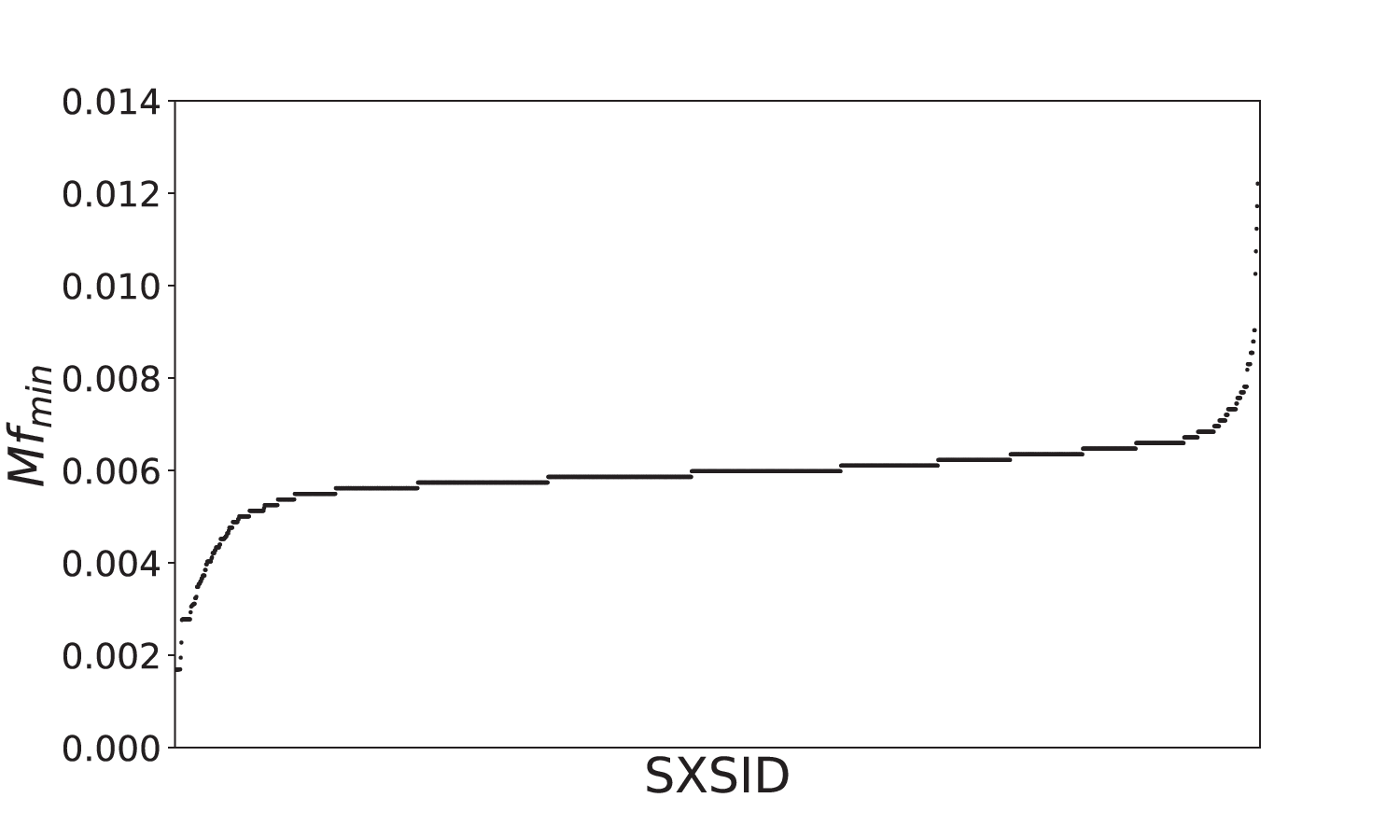}\\
\includegraphics[width=0.45\textwidth]{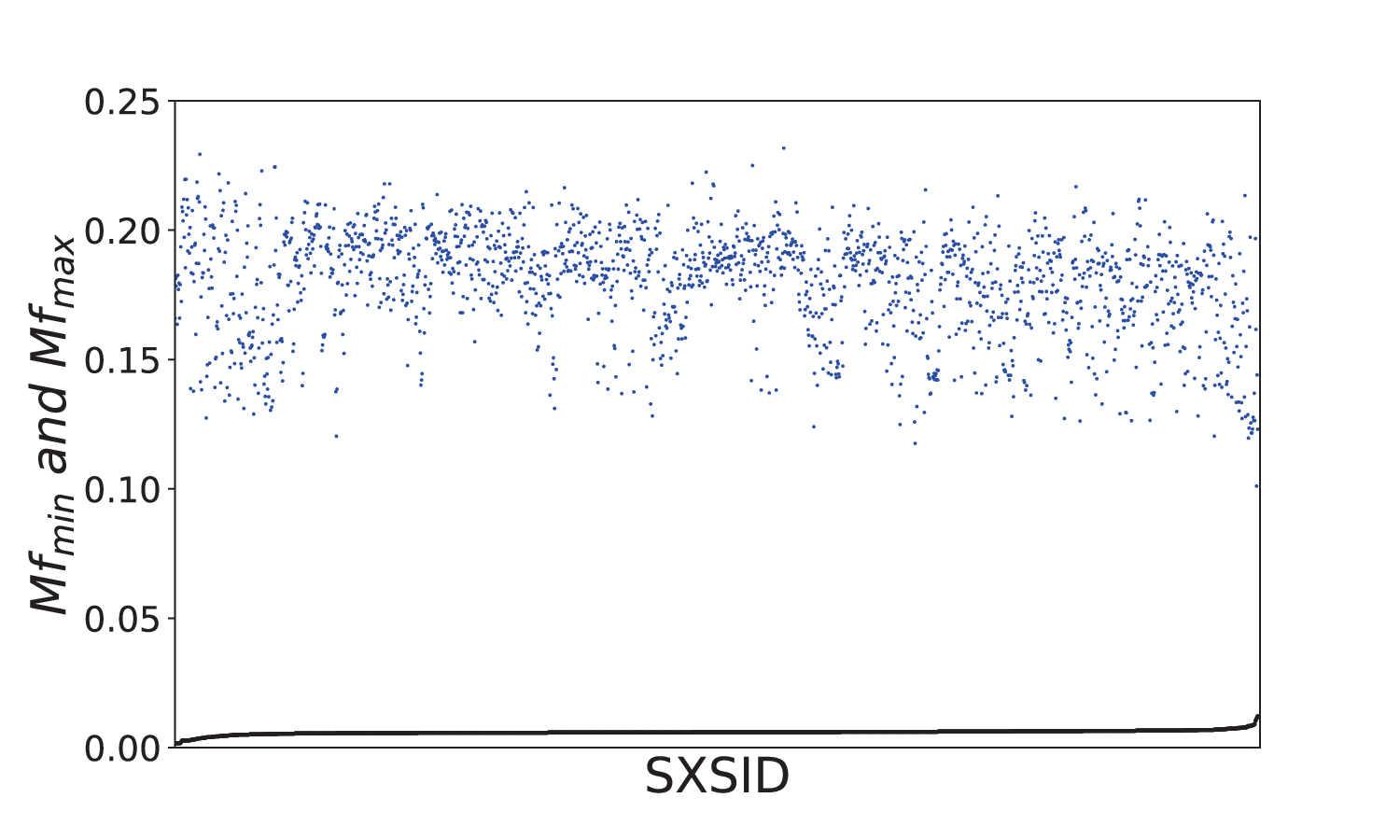}
\end{tabular}
\caption{Frequency lower and upper limit of the 1872 numerical relativity waveforms in SXS catalog. The top panel is the lower limit $Mf_{\rm min}$. The bottom panel shows both the lower limit (black dots) and the upper limit (blue dots).}\label{fig2}
\end{figure}

There are 1872 waveforms in the SXS catalog \cite{SXSBBH} who have more than one resolution result. In Fig.~\ref{fig2}(a) and (b) we plot $Mf_{\rm min}$ and $Mf_{\rm max}$ of these waveforms. Here $M$ means the total mass of the binary system. Different numerical relativity waveforms begin at different frequency corresponding to $Mf_{\rm min}$. $Mf_{\rm min}$ ranges from about 0.002 to 0.012. Most waveforms admit $Mf_{\rm min}\approx0.006$. Lower $Mf_{\rm min}$ means the corresponding binary system begins at larger separation and the waveform is longer. Roughly $Mf_{\rm max}$ falls in the quasi-normal modes stage. The specific value of $Mf_{\rm max}$ depends on the specific numerical simulation. In the viewpoint of the resolution requirement of the binary system in question, if the numerical resolution is higher the value of $Mf_{\rm max}$ is larger. Relatively the numerical setting is random, so the behavior of $Mf_{\rm max}$ shown in Fig.~\ref{fig2}(b) is random.

\begin{figure}
\begin{tabular}{c}
\includegraphics[width=0.45\textwidth]{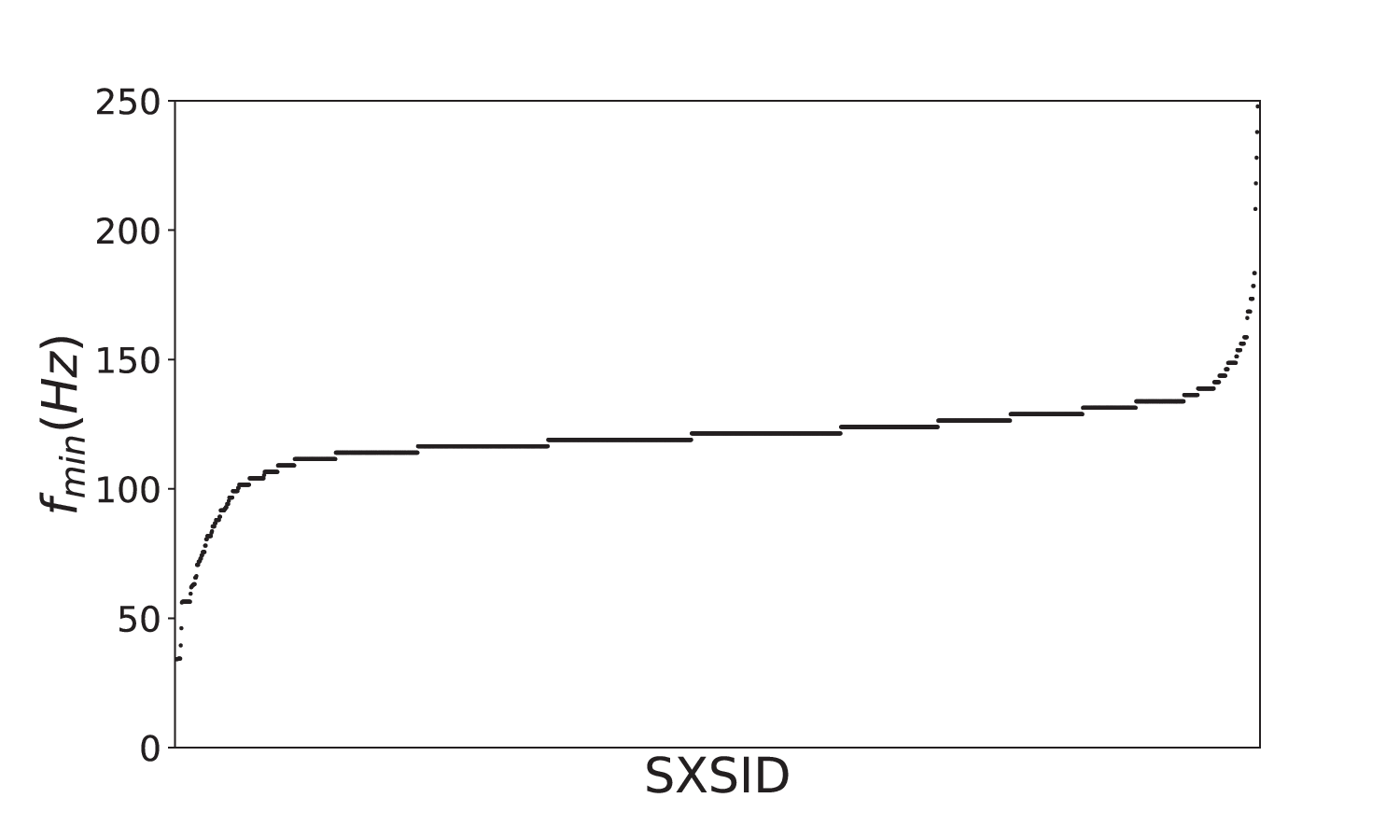}\\
\\
\includegraphics[width=0.45\textwidth]{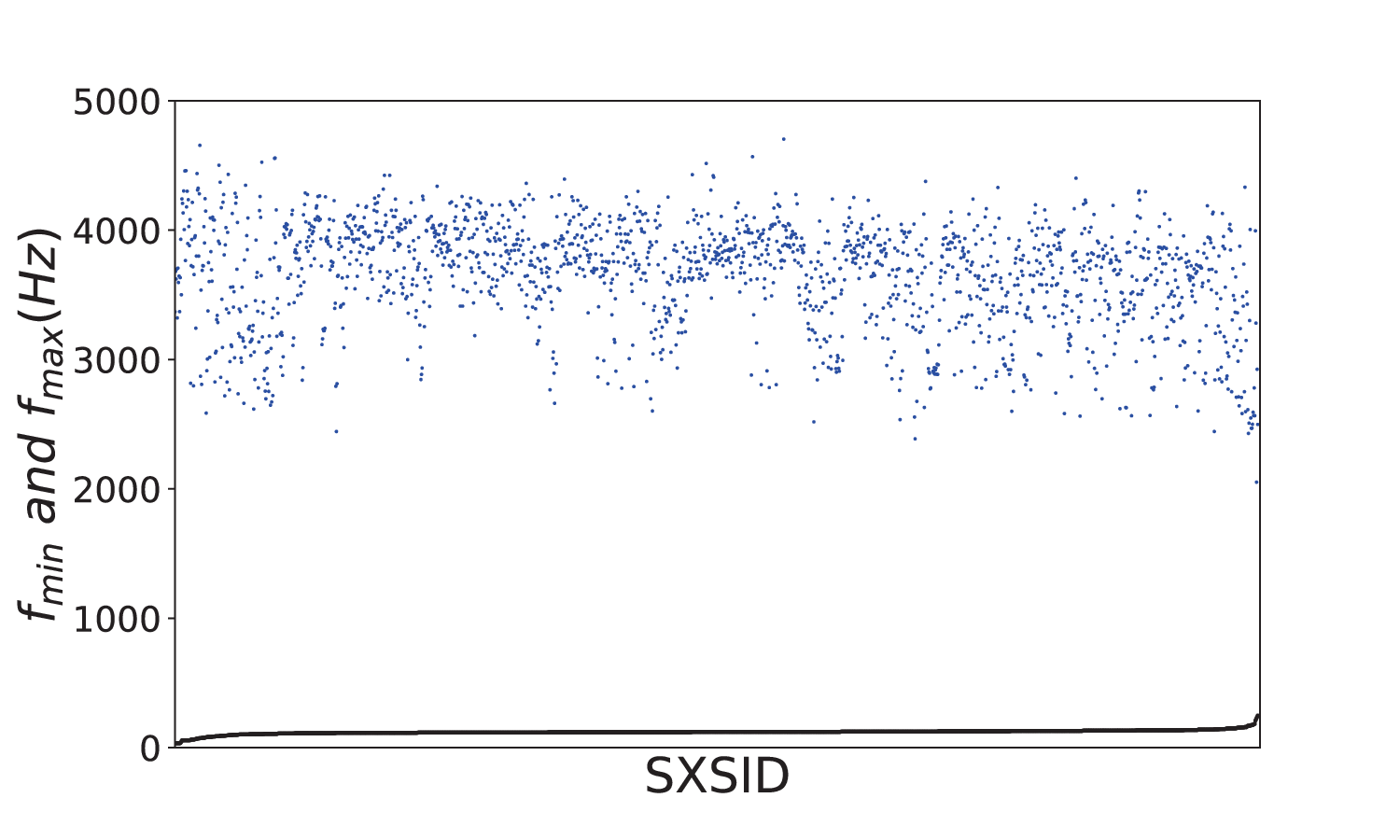}
\end{tabular}
\caption{Trustable frequency range of numerical relativity waveforms of the 1872 numerical relativity waveforms in SXS catalog for $M=10M_\odot$ binary system. The top plot is the lower limit $Mf_{\rm min}$. The bottom plot shows both the lower limit (black dots) and the upper limit (blue dots).}\label{fig3}
\end{figure}

\begin{figure*}[t]
\centering
\begin{tabular}{cc}
\includegraphics[width=0.5\textwidth]{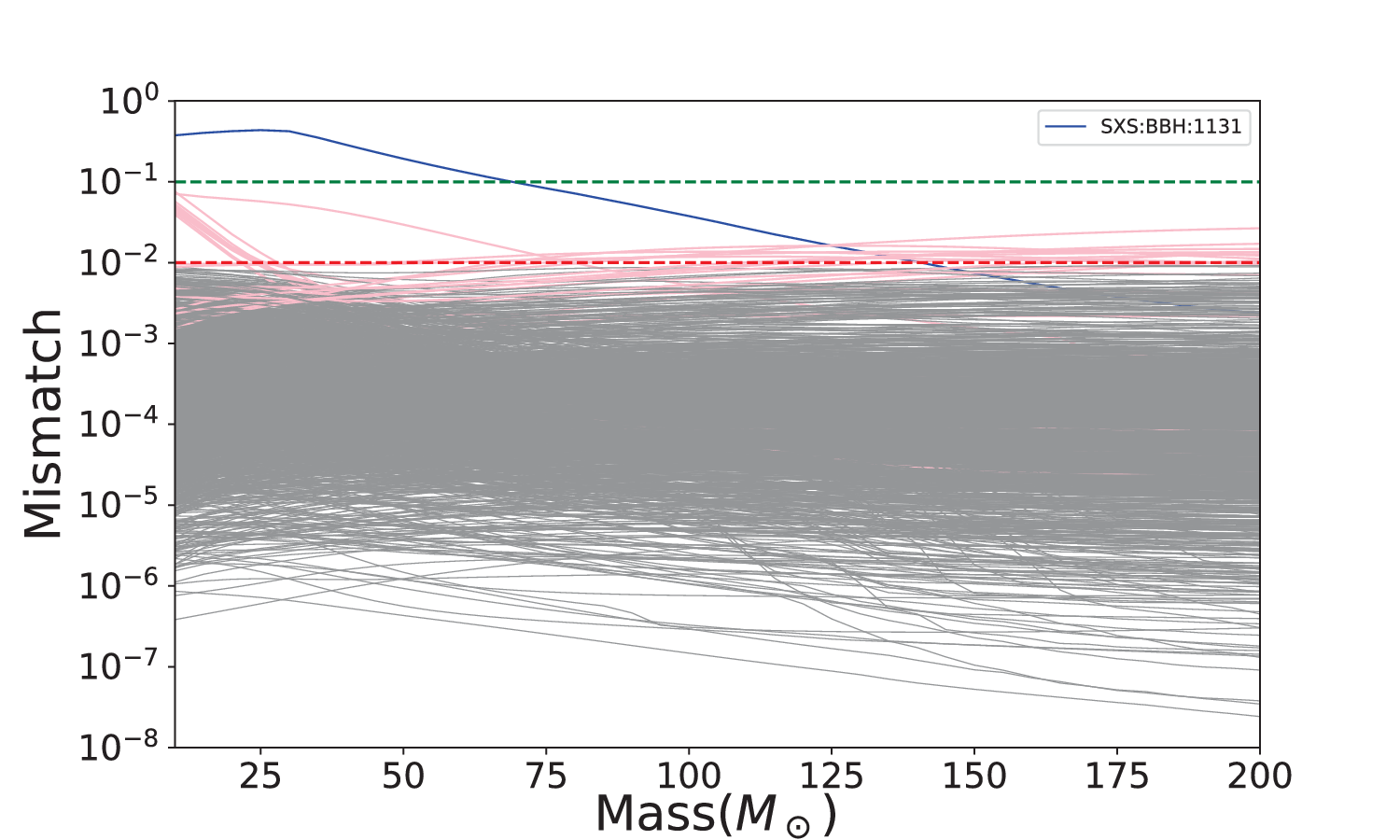}&
\includegraphics[width=0.5\textwidth]{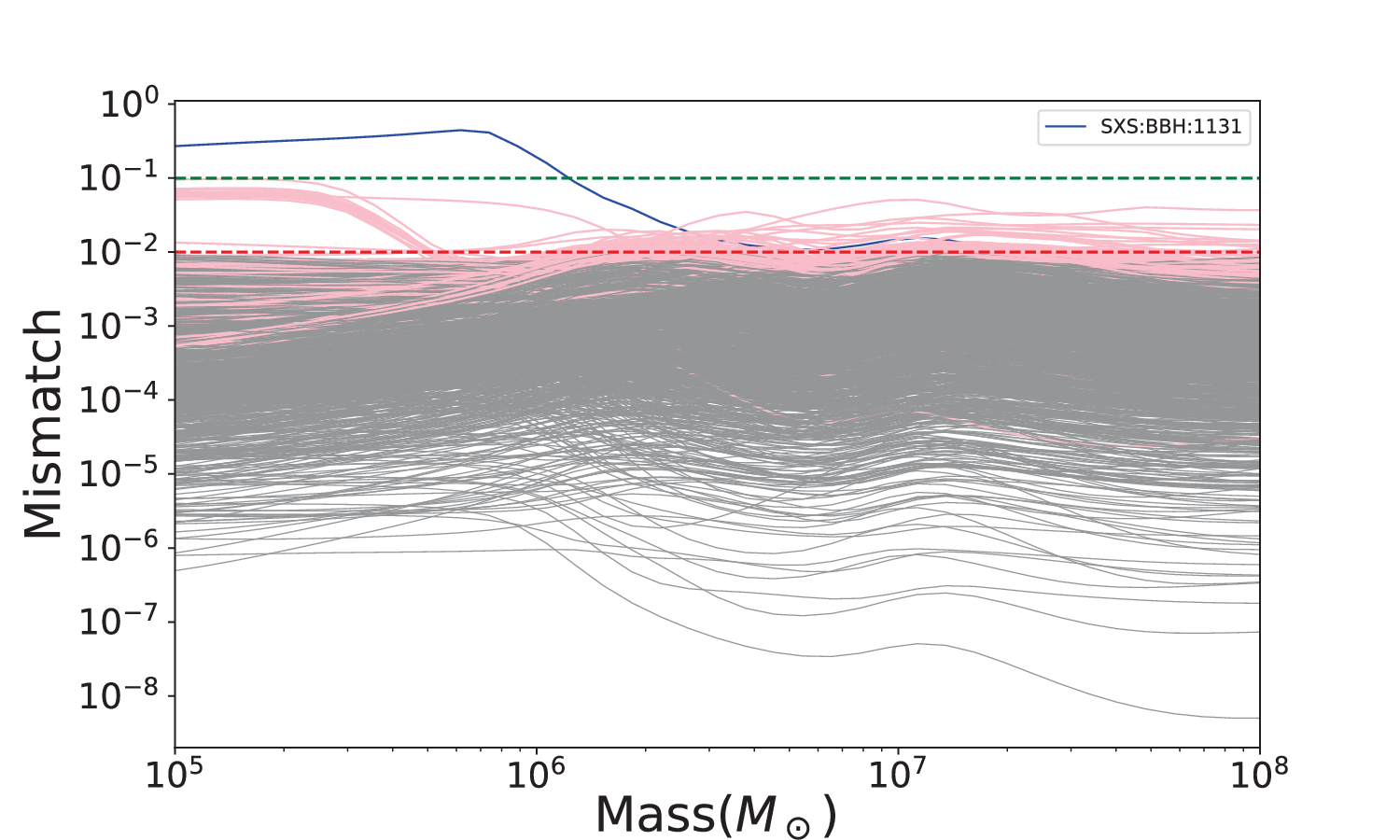}\\
\includegraphics[width=0.5\textwidth]{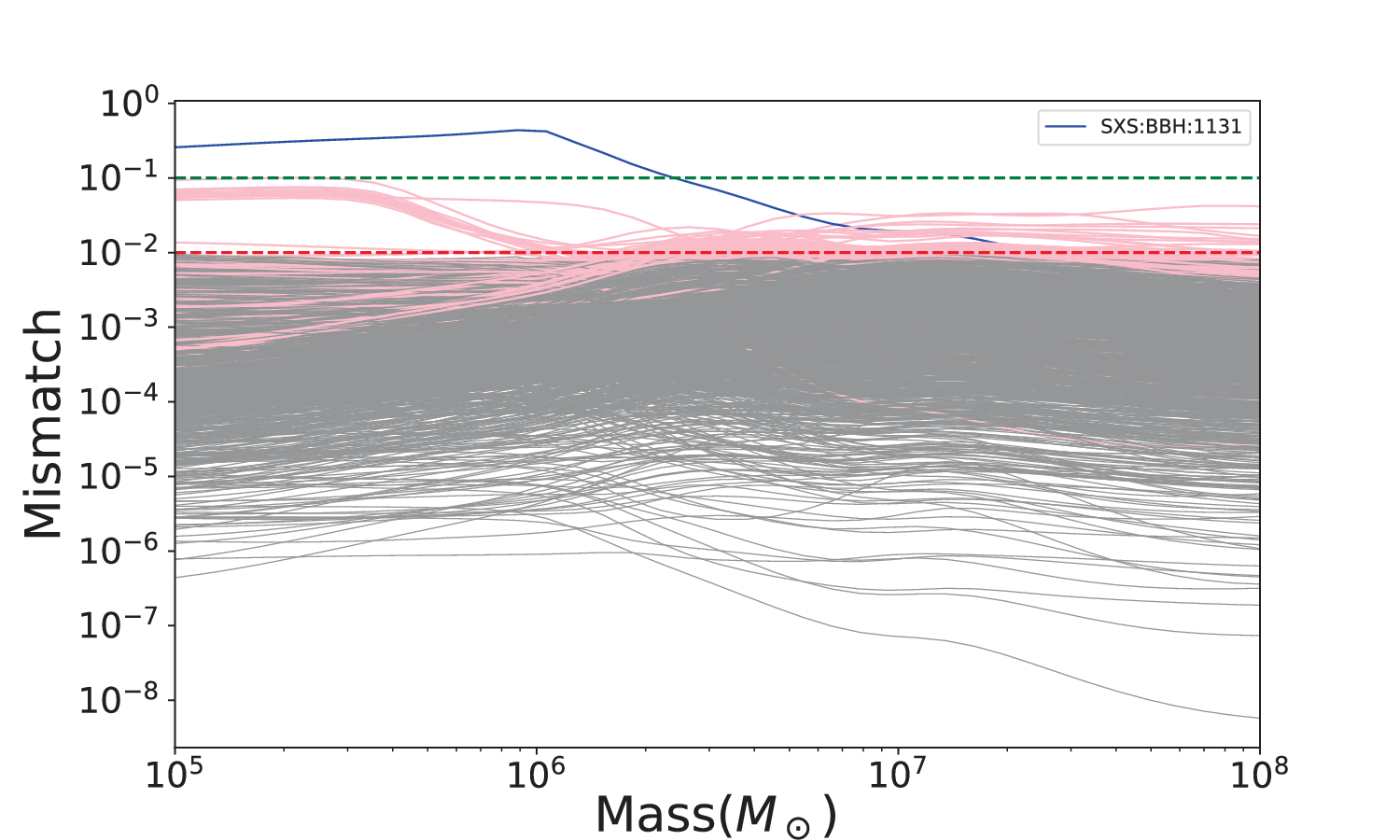}&
\includegraphics[width=0.5\textwidth]{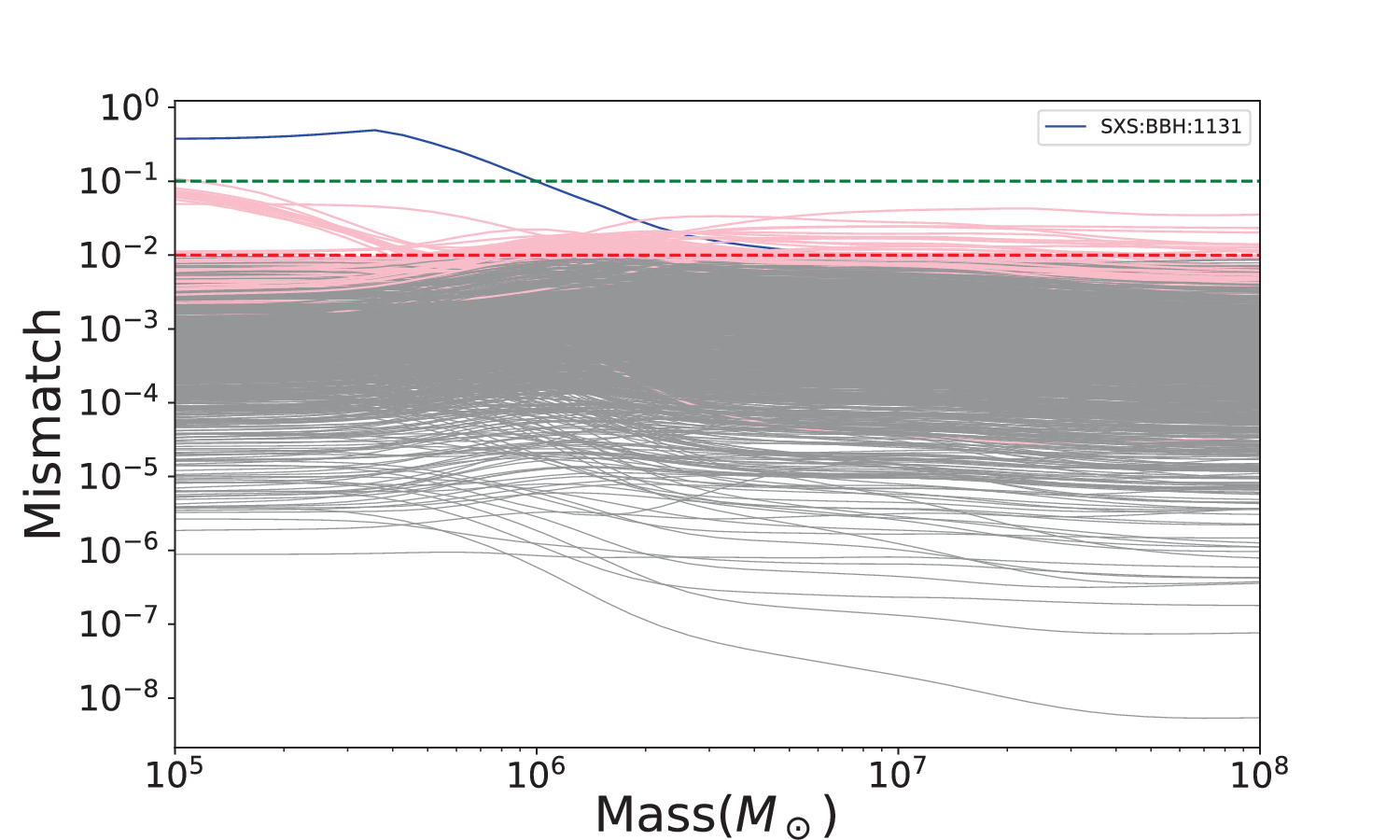}
\end{tabular}
\caption{The mismatch factors between the highest resolution simulation and the second highest resolution simulation. There are in all 1872 SXS waveforms are investigated here. From top to bottom, from left to right the subfigures correspond to LIGO, LISA, Taiji and Tianqin respectively. The blue lines in all of the subfigures correspond to SXS:BBH:1131.}
\label{fig4}
\end{figure*}
\subsection{Accuracy of numerical relativity waveforms with respect to LIGO}
For comparison convenience, we also investigate the accuracy of numerical relativity waveforms with respect to LIGO detectors. Specifically we use the designed sensitivity of advanced LIGO \cite{Sho10}. The frequency band of LIGO is $(10,8192)$Hz.

Note that only the numerical relativity waveform falling in the range $(Mf_{\rm min},Mf_{\rm max})$ is trustable. Considering the source character for LIGO, we investigate $M\in(10,200)M_\odot$. In Fig.~\ref{fig3} we show the trustable frequency range for binary system with total mass $M=10M_\odot$. For other total mass systems we need only rescale the vertical axis proportional to the inverse of the system total mass $1/M$. From Fig.~\ref{fig3}(a) we can see clearly that the numerical relativity simulation can not cover the whole frequency range of LIGO detection. This is due to the well known expensive computational cost of numerical relativity. Consequently numerical relativity only starts near merger. For early inspiral part, people rely on post-Newtonian approximation to construct the waveform template. In the current work, we just care about the accuracy of numerical relativity, so we take the integrand bound in (\ref{equation2}) as
\begin{align}
&f_{\rm low}=\max(10,f_{\rm min}),\\
&f_{\rm up}=\min(8192,f_{\rm max}).
\end{align}

In the first panel of Fig.~\ref{fig4} we plot the mismatch factors
\begin{align}
\mathcal{M}\equiv1-{\rm FF}
\end{align}
with respect to LIGO between the highest resolution simulation and the second highest resolution simulation.
\subsection{Accuracy of numerical relativity waveforms with respect to space-based detectors}
Regarding space-based detectors, we consider LISA \cite{2023LRR....26....2A,PhysRevLett.118.171101}, Taiji \cite{Ruan2020} and Tianqin \cite{luo2016tianqin,Luo_2020} as examples. We do not involve realistic response functions as \cite{PhysRevD.102.124037}, instead we use sky averaged sensitivity \cite{Robson_2019} to do the estimation.

Specifically we use the following approximated sensitivity for space based gravitational wave detectors (Eq.~(13) of \cite{Robson_2019})
\begin{align}
S_n(f)&=\frac{10}{3L^2}\left(P_{\rm OMS}+2(1+\cos^2(f/f_*))\frac{P_{\rm acc}}{(2\pi f)^4}\right)\times\nonumber\\
&\left(1+\frac{6}{10}\left(\frac{f}{f_*}\right)^2\right),\\
f_*&=c/(2\pi L).
\end{align}
For LISA \cite{Robson_2019} we have
\begin{align}
P_{\rm OMS}&=(1.5\times10^{-11}{\rm m})^2{\rm Hz}^{-1},\\
P_{\rm acc}&=(3\times10^{-15}{\rm ms}^{-2})^2\left(1+\left(\frac{4\times10^{-4}{\rm Hz}}{f}\right)^2\right){\rm Hz}^{-1},\\
L&=2.5\times10^9{\rm m}.
\end{align}
For Taiji \cite{2020IJMPA..3550075R} we have
\begin{align}
P_{\rm OMS}&=(8\times10^{-12}{\rm m})^2{\rm Hz}^{-1},\\
P_{\rm acc}&=(3\times10^{-15}{\rm ms}^{-2})^2\left(1+\left(\frac{4\times10^{-4}{\rm Hz}}{f}\right)^2\right){\rm Hz}^{-1},\\
L&=3\times10^9{\rm m}.
\end{align}
For Tianqin we have \cite{luo2016tianqin}
\begin{align}
P_{\rm OMS}&=(1\times10^{-12}{\rm m})^2{\rm Hz}^{-1},\\
P_{\rm acc}&=(1\times10^{-15}{\rm ms}^{-2})^2\left(1+\left(\frac{1\times10^{-4}{\rm Hz}}{f}\right)^2\right){\rm Hz}^{-1},\\
L&=\sqrt{3}\times10^8{\rm m}.
\end{align}

\begin{figure}[h]
\centering
\begin{tabular}{c}
\includegraphics[width=0.5\textwidth]{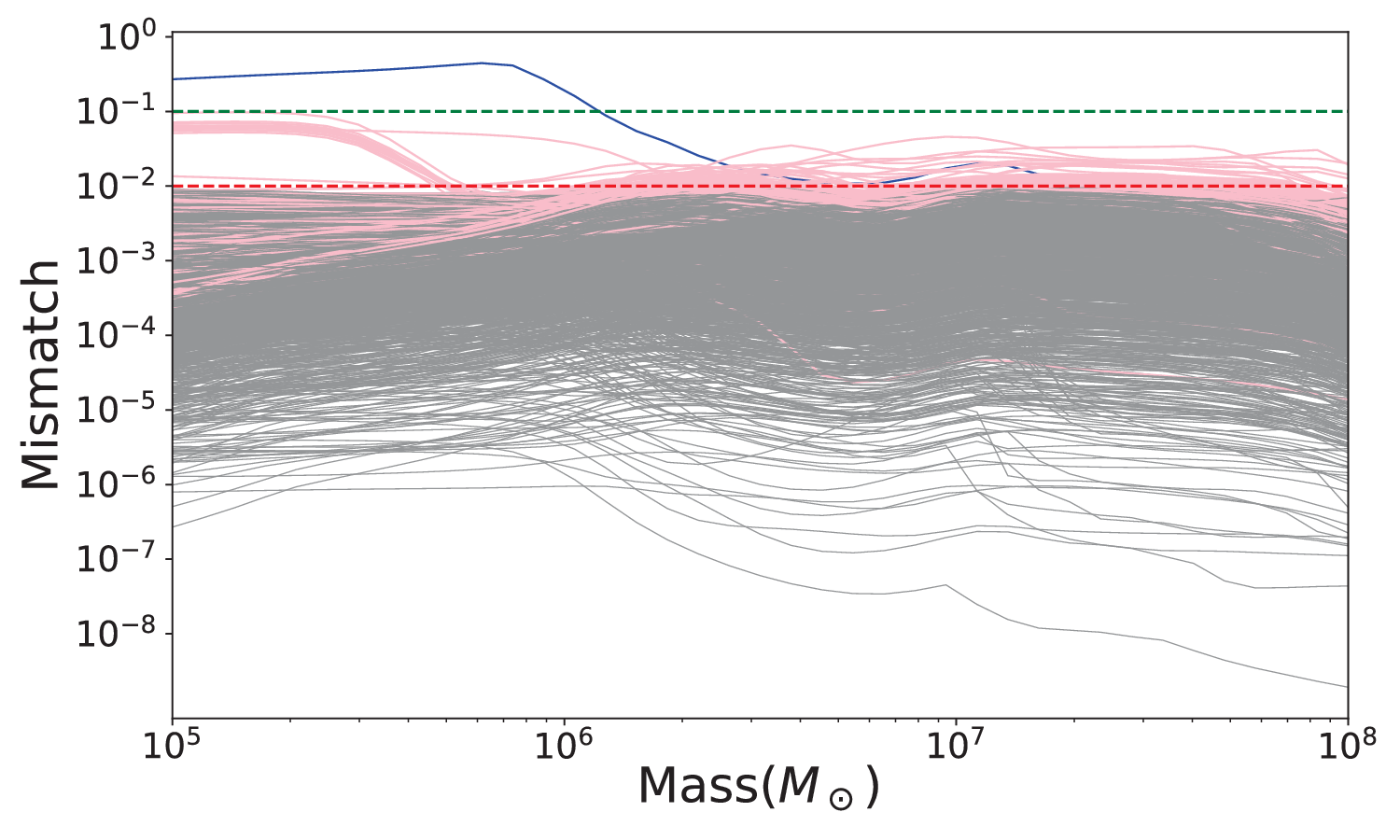}
\end{tabular}
\caption{Similar to Fig.~\ref{fig4} but with detector frequency range $(10^{-4},0.1)$Hz in stead of $(10^{-5},1)$Hz. This plot is for LISA.}
\label{fig5}
\end{figure}

Besides the instrument noise mentioned above, there is more confusion noise due to the galaxy binaries which can be approximated as \cite{Cornish_2017}
\begin{align}
    S_{c}(f)=&A f^{-7 / 3} \mathrm{e}^{-f^{\alpha}+\beta f \sin (\kappa f)}\times\nonumber\\
    &\left[1+\tanh \left(\gamma\left(f_{k}-f\right)\right)\right] \mathrm{Hz}^{-1}\\
    A&=9\times10^{-45},\\
    \alpha&=0.133,\\
    \beta&=243,\\
    \kappa&=482,\\
    \gamma&=917,\\
    f_k&=0.00258.
\end{align}
Note that parameters $\alpha,\beta,\kappa,\gamma$ depend on observation time. The values listed above correspond to observation time half year. The overall noise sensitivity of space-based detectors can be estimated as
\begin{align}
S=S_n+S_c.
\end{align}

Due to the similar reason for LIGO, we take the integrand bound in (\ref{equation2}) as
\begin{align}
&f_{\rm low}=\max(10^{-5},f_{\rm min}),\label{eq1}\\
&f_{\rm up}=\min(1,f_{\rm max}),\label{eq2}
\end{align}
for space-based detectors. 

\clearpage
\begin{longtable*}{cccccccccc}
    \caption{Less accurate ($\mathcal{M}>1\%$) NR simulations found in this work. Here, we list the parameters for each simulation, including mass ratio $q$, lowest frequency $Mf_{\rm min}$, highest frequency $Mf_{\rm max}$, and initial spin configuration. Additionally, we provide the maximum mismatch between the highest resolution simulation and the second highest resolution simulation $\max\mathcal{M}$. Here max means the maximum value in the total mass range shown in Fig.~\ref{fig4}. The subscriptions `LIGO', `LISA', `Taiji' and `Tianqin' are for corresponding detectors.}\label{table1} \\
    \hline
    \hline
    SXS ID & $q$ & $\pmb{\chi}_1$ & $\pmb{\chi}_2$ & $Mf_{\rm min}$ & $Mf_{\rm max}$ & $\max\mathcal{M}_{\rm LIGO}$ & $\max\mathcal{M}_{\rm LISA}$ & $\max\mathcal{M}_{\rm Taiji}$ & $\max\mathcal{M}_{\rm Tianqin}$\\
    \hline
    \endfirsthead
    \multicolumn{10}{c}{\bfseries\small \tablename\ \thetable\ {continue}}\\
    \hline
    \hline
    SXS ID & $q$ & $\pmb{\chi}_1$ & $\pmb{\chi}_2$ & $Mf_{\rm min}$ & $Mf_{\rm max}$ & $\max\mathcal{M}_{\rm LIGO}$ & $\max\mathcal{M}_{\rm LISA}$ & $\max\mathcal{M}_{\rm Taiji}$ & $\max\mathcal{M}_{\rm Tianqin}$\\
    \hline
    \endhead
    \hline
    \multicolumn{10}{r@{}}{\textit{to next page}}
    \endfoot
    \hline
    \hline
    \endlastfoot    %
    \input{sxsprectable.tex}
\end{longtable*}

NR waveforms have a critical limitation that they are some short (due to the computational cost) and mainly focus on the merger phase. Especially for the gravitational waves emitted by supermassive black hole binaries, the majority of the evolution occurs in the inspiral phase. Therefore, simply calculating the accuracy of NR waveforms will lose the important inspiral phase, which will affect the results of the accuracy of the waveforms. In future work, we plan to use the PN(Post-Newtonian)-NR waveform models including SEOBNR, SEOBNRE and others to investigate the waveform template accuracy for space-based detectors.

The corresponding mismatch factors between the highest resolution simulation and the second highest resolution simulation for LISA, Taiji and Tianqin are shown in Fig.~\ref{fig4}. Similar to the situation for LIGO, most NR simulations admit accuracy better than 99\%. A few NR simulations have less accuracy. We list these less accurate simulations in Tab.~\ref{table1}.

From Fig.~\ref{fig4}, we can see SXS:BBH:1131 has very large mismatch factor. This means ones must take caution when using SXS:BBH:1131 result. For other simulations listed in Tab.~\ref{table1}, ones also have to note the specific accuracy requirement when using those simulation results.

For all lines of Fig.~\ref{fig4}, there is a typical behavior that the line increases along with the black hole mass and then decreases. We can understand this fact as follows. Due to the numerical error accumulation, the merger part of the waveform corresponds to the least accurate part of the waveform. Due to the resolution requirement of the simulation, the merger part is also the least accurate part of the waveform. In the frequency domain, when the black hole mass increases, the merger part moves from right to left. Note that the most sensitive range of the detector locates at the center. For relative small mass BBHs, the merger part waveform locates at the right side of the aforementioned sensitive frequency range. When black hole mass increases, the merger part falls into the sensitive frequency range. Consequently the mismatch factor increases. When the black hole mass increases more, the merger part waveform leaves the sensitive frequency range. So the mismatch factor decreases consequently.

Comparing to the result for LIGO, we find that the numerical relativity accuracy for space-based detectors is comparable to that for ground-based detectors. That is to say if the accuracy requirement is similar to that of LIGO, the current numerical relativity simulation results can satisfy the need of space-based detectors.

Considering that the frequency range of space-based detector may not reach $(10^{-5},1)$Hz, we have also calculated the mismatch factor with replacing (\ref{eq1}) and (\ref{eq2}) with
\begin{align}
&f_{\rm low}=\max(10^{-4},f_{\rm min}),\\
&f_{\rm up}=\min(0.1,f_{\rm max}).
\end{align}
The results are almost the same as Fig.~\ref{fig4}. Since the results for LISA, Taiji and Tianqin are similar to each other, we only plot LISA as the example in Fig.~\ref{fig5}.

\begin{figure*}[t]
\centering
\begin{tabular}{cc}
\includegraphics[width=0.5\textwidth]{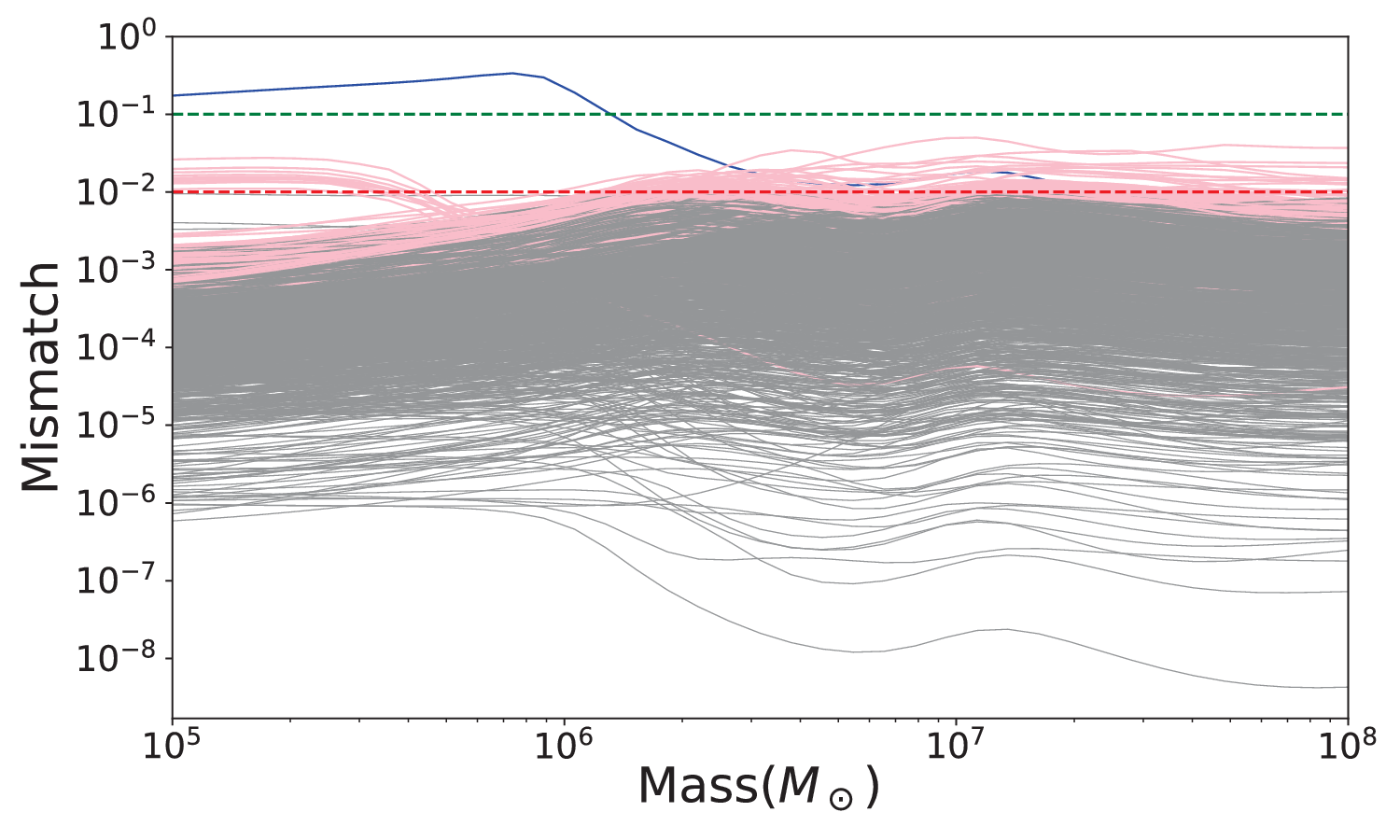}&
\includegraphics[width=0.5\textwidth]{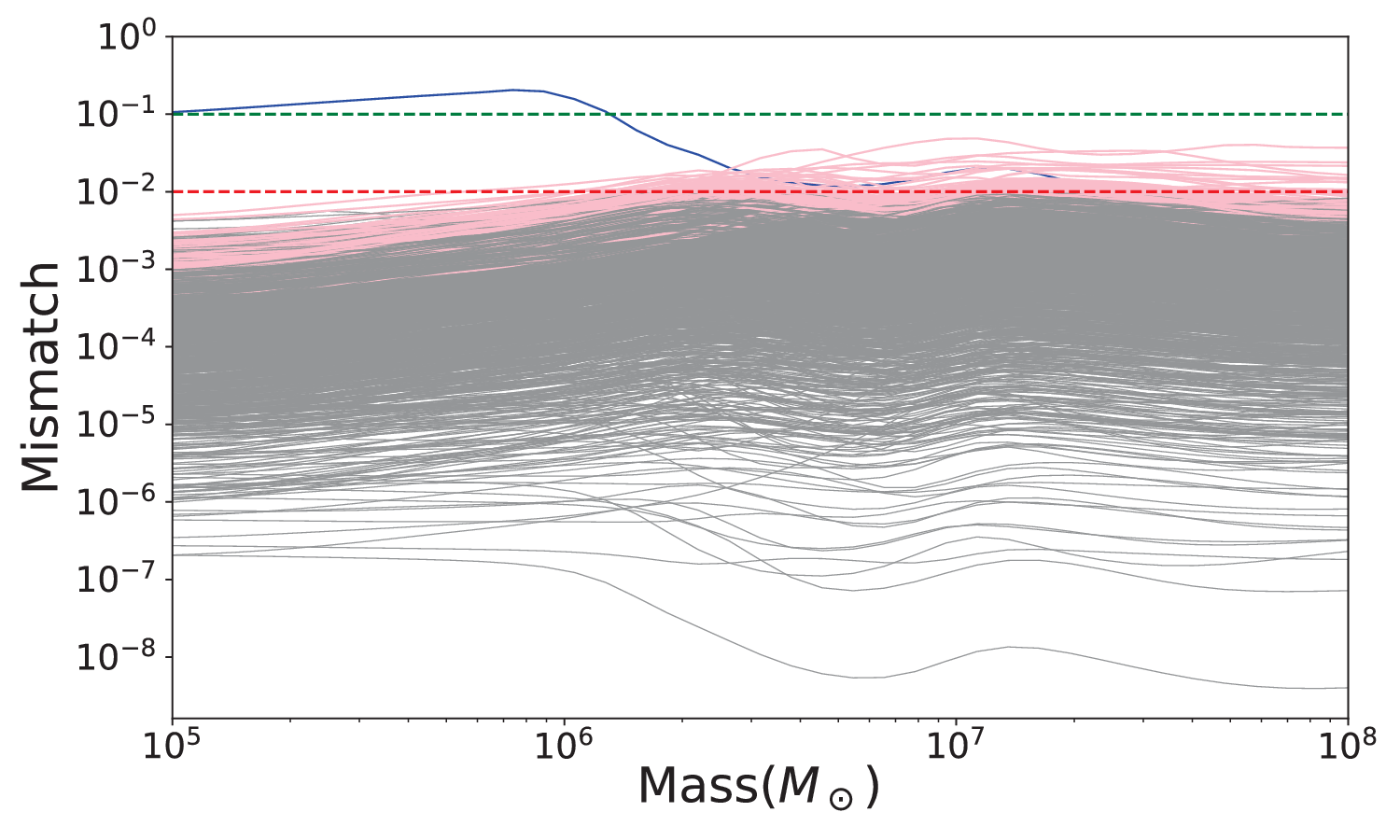}
\end{tabular}
\caption{Similar to Fig.~\ref{fig4} but with $f_{\rm low}=\max(10^{-5},1.2f_{\rm min})$Hz (left panel) and $f_{\rm low}=\max(10^{-5},1.5f_{\rm min})$Hz (right panel) instead of $f_{\rm low}=\max(10^{-5},f_{\rm min})$Hz. Like Fig.~\ref{fig5}, we again use LISA as the example.}
\label{fig6}
\end{figure*}

\begin{figure*}[ht]
\centering
\begin{tabular}{cc}
\includegraphics[width=0.5\textwidth]{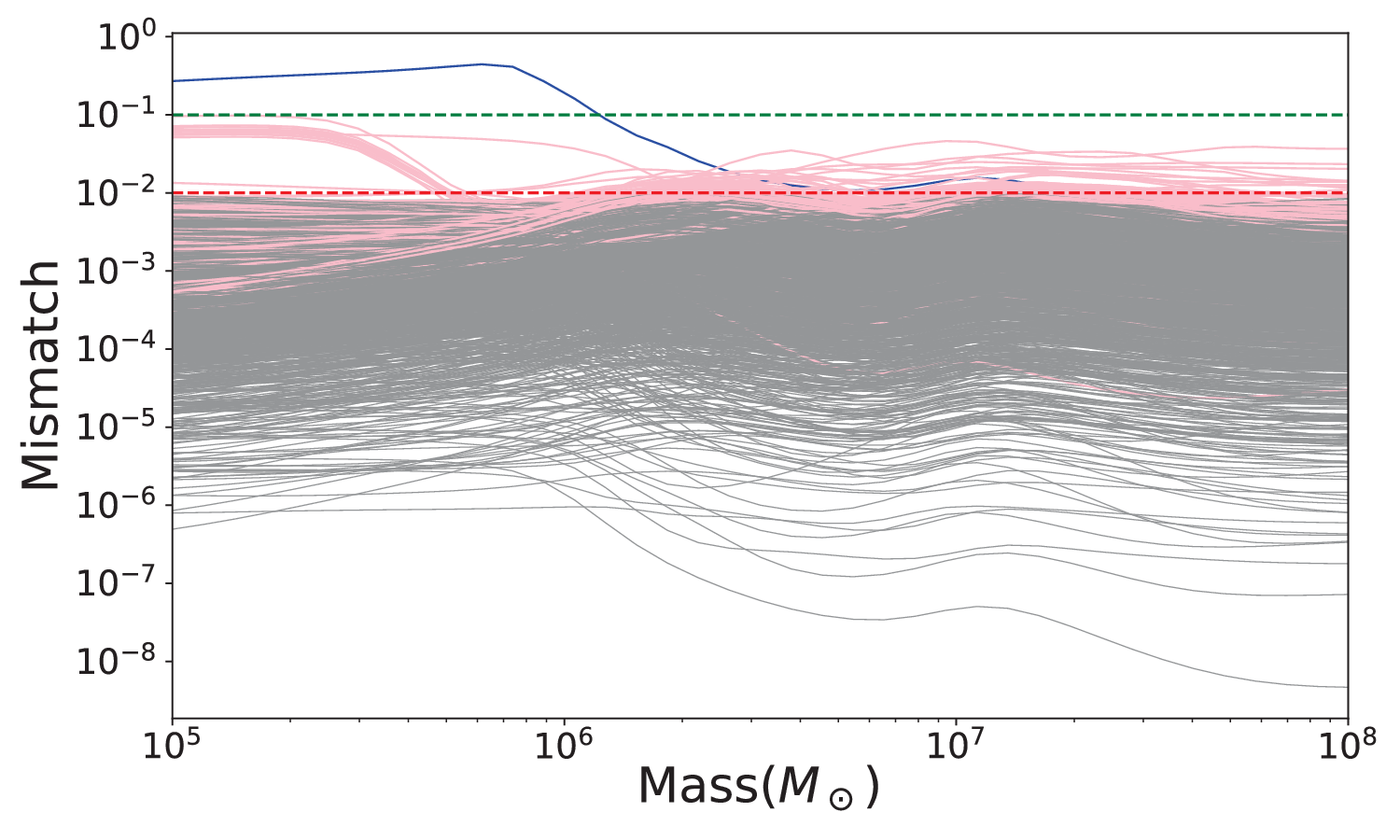}&
\includegraphics[width=0.5\textwidth]{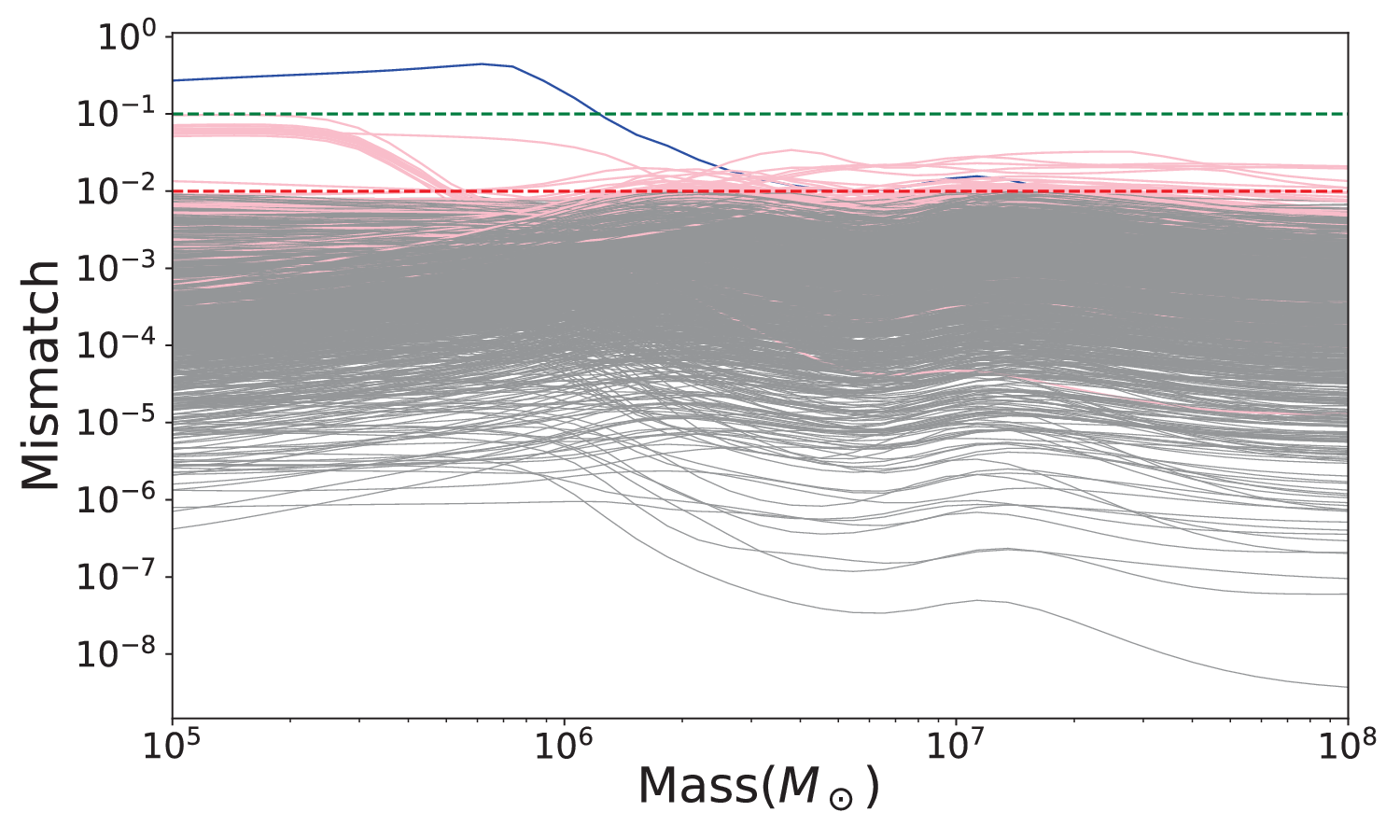}
\end{tabular}
\caption{Similar to Fig.~\ref{fig4} but with $f_{\rm up}=\min(1,0.8f_{\rm max})$Hz (left panel) and $f_{\rm up}=\min(1,0.5f_{\rm max})$Hz (right panel) instead of $f_{\rm up}=\min(1,f_{\rm max})$Hz. Like Fig.~\ref{fig5}, we again use LISA as the example.}
\label{fig7}
\end{figure*}

\begin{figure}[h]
\centering
\begin{tabular}{cc}
\includegraphics[width=0.5\textwidth]{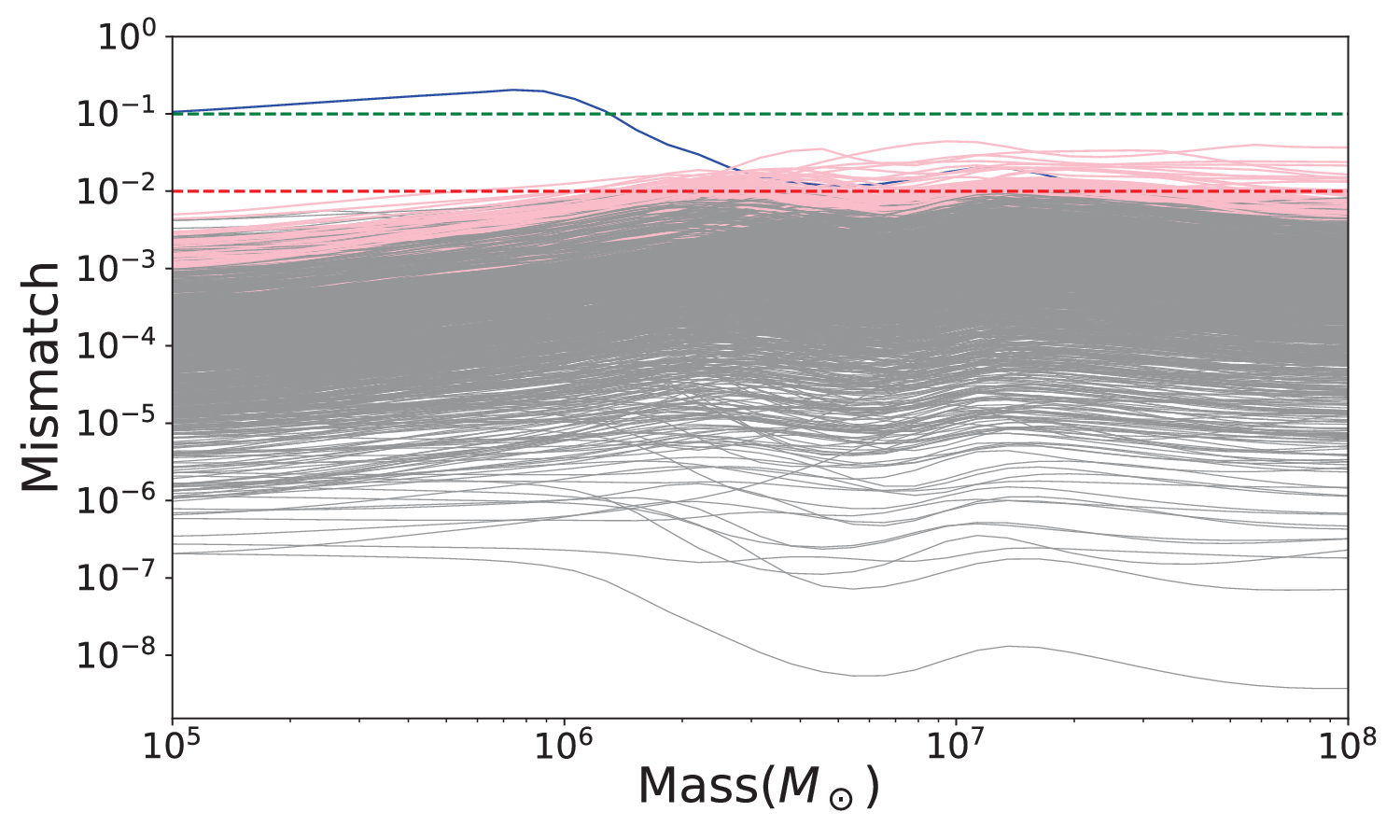}
\end{tabular}
\caption{Similar to Fig.~\ref{fig4} but with $f_{\rm low}=\max(10^{-5},1.5f_{\rm min})$Hz and $f_{\rm up}=\min(1,0.8f_{\rm max})$Hz frequency choice instead of (\ref{eq1}) and (\ref{eq2}). Like Fig.~\ref{fig5}, we again use LISA as the example.}
\label{fig8}
\end{figure}
The frequency range of numerical relativity waveform shown in Fig.~\ref{fig1} is the most optimal one. We can see clear unphysical oscillation near the low frequency $f_{\rm min}$. In order to check the influence of such frequency range choice, we have also considered
\begin{align}
&f_{\rm low}=\max(10^{-5},1.2f_{\rm min}),\\
&f_{\rm up}=\min(1,f_{\rm max}),
\end{align}
and
\begin{align}
&f_{\rm low}=\max(10^{-5},1.5f_{\rm min}),\\
&f_{\rm up}=\min(1,f_{\rm max}).
\end{align}
Similar to Fig.~\ref{fig5}, we once again use LISA as example and plot the results in Fig.~\ref{fig6} for these two frequency range choices. As ones expected, when we consider shorter inspiral part, the waveform accuracy becomes higher. So we can see several lines above $10^{-2}$ in the left panel of Fig.~\ref{fig6} fall down below $10^{-2}$ in the right panel.

Regarding to high frequency side, we check how the cutting frequency affects the waveform accuracy. For comparison we have compared the results plotted in Fig.~\ref{fig4} to frequency choices
\begin{align}
&f_{\rm low}=\max(10^{-5},f_{\rm min}),\\
&f_{\rm up}=\min(1,0.8f_{\rm max}),
\end{align}
and
\begin{align}
&f_{\rm low}=\max(10^{-5},f_{\rm min}),\\
&f_{\rm up}=\min(1,0.5f_{\rm max}).
\end{align}
The result is shown in Fig.~\ref{fig7}. As ones expected, the high frequency side affects large black hole mass systems more. But in all, the influence is small.

And more we have also considered conservative frequency range choice on both low and high frequency side
\begin{align}
&f_{\rm low}=\max(10^{-5},1.5f_{\rm min}),\\
&f_{\rm up}=\min(1,0.8f_{\rm max}).
\end{align}
The result is plotted in Fig.~\ref{fig8}. In a short summary, the different frequency choices roughly result in similar waveform accuracy.

\section{Summary and conclusion}
One of the most challenging and fascinating problems in gravitational physics is to understand the dynamics of binary black hole mergers in the strong-field regime. In this regime, the components of the binary move at relativistic speeds and the spacetime curvature becomes highly nonlinear, making analytical approximations inadequate. The only reliable way to obtain precise solutions to Einstein's field equations in this regime is to use numerical relativity, which involves solving the full nonlinear equations on high-performance computers. This breakthrough was achieved in 2005 after decades of efforts \cite{Zhao2020}.

Numerical relativity simulations of binary black hole mergers are essential for modeling the gravitational wave signals emitted by these systems during their late inspiral, merger, and ringdown phases. These signals are used to infer the properties of the source systems and to test general relativity in extreme conditions. All binary black hole detections made by LIGO and Virgo have been analyzed using waveform models that incorporate numerical relativity data. The most prominent examples of these models are the effective-one-body and phenomenological waveform models. Numerical relativity also plays a key role in validating these models and testing their accuracy and robustness. Moreover, numerical relativity waveforms can be directly used for parameter estimation, template bank construction, and waveform family development without intermediate analytical steps, using techniques such as reduced order modeling.

Several coordinated efforts have been undertaken to produce numerical relativity simulations of binary black hole mergers for gravitational wave applications. These include the Numerical Injection Analysis (NINJA) project \cite{2014CQGra..31k5004A}, the collaboration between Numerical Relativity and Analytical Relativity (NRAR), and the waveform catalogs released by the SXS collaboration and Georgia Tech.

In this work, we use numerical simulations of binary black hole mergers performed by the SXS Collaboration using the Spectral Einstein Code (SpEC). The SXS catalog has been used to construct SEOBNRE waveform model \cite{PhysRevD.96.044028,PhysRevD.101.044049,2022CQGra..39c5009L,IJMPD.32.2350015} and other waveform models. The accuracy of numerical relativity waveform is very important to gravitational wave astronomy study.

In previous works, the accuracy issue of numerical relativity waveform has been well studied for ground based detectors. In the current paper, we focus on space-based detectors. We have systematically investigated the effect of the waveform frequency range, the detector sensitivity detail, the BBH's black hole mass and others on the waveform accuracy issue.

Each waveform of SXS catalog has been investigated. Special attention is payed to matching factor calculation between highest and second highest resolution used in the numerical simulations. Our calculation results indicate that the numerical relativity waveforms are as accurate as 99\% with respect to space-based detectors including LISA, Taiji and Tianqin. Such accuracy level is comparable to the one with respect to LIGO. If only the accuracy requirement for space-based detectors is similar to that of ground-based ones, the current numerical relativity waveforms are valid for waveform modelling.

\section*{Acknowledgments}
This work was supported in part by the National Key Research and Development Program of China Grant No. 2021YFC2203001 and in part by the NSFC (No.~11920101003, No.~12021003 and No.~12005016). Z. Cao was supported by ``the Interdiscipline Research Funds of Beijing Normal University" and CAS Project for Young Scientists in Basic Research YSBR-006.

\bibliographystyle{unsrt}
\bibliography{refs}

\end{document}

%% file: sxsprectable.tex
1415 & 1.50 & (0.00,0.00,0.50) & (0.00,-0.00,0.50) & 0.0017 & 0.1789 & 0.0756 & 0.0978 & 0.1005 & 0.1077 \\
0627 & 1.91 & (-0.51,0.44,-0.35) & (0.19,-0.01,-0.06) & 0.0083 & 0.1626 & 0.0709 & 0.0650 & 0.0660 & 0.0491 \\
1413 & 1.41 & (-0.00,-0.00,0.50) & (-0.00,-0.00,0.40) & 0.0017 & 0.1660 & 0.0570 & 0.0732 & 0.0755 & 0.0805 \\
1414 & 1.83 & (-0.00,-0.00,-0.50) & (0.00,-0.00,0.40) & 0.0017 & 0.1636 & 0.0527 & 0.0695 & 0.0715 & 0.0749 \\
1390 & 1.42 & (0.15,0.44,-0.16) & (-0.02,0.34,0.10) & 0.0017 & 0.1659 & 0.0510 & 0.0648 & 0.0672 & 0.0714 \\
1393 & 1.79 & (-0.37,-0.33,-0.00) & (-0.27,-0.39,0.11) & 0.0017 & 0.1784 & 0.0490 & 0.0647 & 0.0667 & 0.0694 \\
1392 & 1.51 & (-0.40,0.23,0.17) & (0.35,-0.13,-0.25) & 0.0017 & 0.1795 & 0.0478 & 0.0625 & 0.0642 & 0.0679 \\
1389 & 1.63 & (-0.29,0.20,-0.30) & (-0.01,0.42,0.16) & 0.0017 & 0.1771 & 0.0460 & 0.0594 & 0.0613 & 0.0651 \\
1391 & 1.83 & (-0.15,0.29,-0.33) & (-0.33,-0.29,-0.03) & 0.0017 & 0.1813 & 0.0440 & 0.0566 & 0.0583 & 0.0619 \\
1412 & 1.63 & (-0.00,-0.00,0.40) & (-0.00,0.00,-0.30) & 0.0017 & 0.1822 & 0.0421 & 0.0562 & 0.0578 & 0.0606 \\
1416 & 1.78 & (0.00,-0.00,-0.40) & (-0.00,0.00,-0.40) & 0.0017 & 0.1825 & 0.0392 & 0.0524 & 0.0539 & 0.0561 \\
1926 & 4.00 & (0.76,0.26,0.04) & (0.00,-0.14,0.79) & 0.0065 & 0.1943 & 0.0266 & 0.0350 & 0.0337 & 0.0335 \\
2000 & 4.00 & (-0.40,0.69,0.08) & (0.45,0.65,-0.11) & 0.0066 & 0.1902 & 0.0171 & 0.0206 & 0.0204 & 0.0204 \\
1992 & 4.00 & (-0.61,0.07,-0.51) & (-0.27,0.75,-0.05) & 0.0062 & 0.1733 & 0.0162 & 0.0192 & 0.0195 & 0.0194 \\
2044 & 4.00 & (0.74,-0.29,0.11) & (0.14,-0.60,0.52) & 0.0067 & 0.1890 & 0.0148 & 0.0151 & 0.0149 & 0.0140 \\
1991 & 4.00 & (-0.26,-0.51,-0.56) & (-0.07,0.06,0.79) & 0.0061 & 0.1724 & 0.0136 & 0.0201 & 0.0219 & 0.0221 \\
2038 & 4.00 & (-0.80,-0.05,0.05) & (-0.01,-0.08,-0.39) & 0.0065 & 0.1920 & 0.0135 & 0.0249 & 0.0244 & 0.0246 \\
2054 & 4.00 & (0.66,-0.45,0.08) & (0.38,-0.31,0.63) & 0.0065 & 0.1887 & 0.0135 & 0.0199 & 0.0190 & 0.0192 \\
2074 & 4.00 & (-0.66,0.44,0.07) & (-0.74,0.28,0.10) & 0.0066 & 0.1548 & 0.0127 & 0.0214 & 0.0210 & 0.0211 \\
1987 & 4.00 & (0.38,0.43,-0.55) & (0.54,0.58,0.04) & 0.0062 & 0.1659 & 0.0119 & 0.0160 & 0.0177 & 0.0182 \\
1110 & 7.00 & (-0.00,-0.00,0.00) & (-0.00,-0.00,-0.00) & 0.0023 & 0.1724 & 0.0106 & 0.0510 & 0.0423 & 0.0427 \\
1928 & 4.00 & (-0.33,0.72,0.07) & (0.62,0.48,-0.13) & 0.0065 & 0.1898 & 0.0104 & 0.0114 & 0.0115 & 0.0113 \\
1978 & 4.00 & (0.50,0.26,0.57) & (-0.77,-0.20,0.03) & 0.0070 & 0.1438 & 0.0092 & 0.0171 & 0.0168 & 0.0179 \\
1135 & 1.00 & (-0.00,-0.00,-0.44) & (-0.00,0.00,-0.44) & 0.0073 & 0.1355 & 0.0089 & 0.0134 & 0.0137 & 0.0097 \\
1623 & 3.93 & (0.02,0.55,0.43) & (-0.56,-0.33,-0.45) & 0.0066 & 0.1863 & 0.0089 & 0.0187 & 0.0184 & 0.0176 \\
1993 & 4.00 & (-0.06,-0.58,-0.54) & (-0.24,-0.76,0.02) & 0.0062 & 0.1439 & 0.0087 & 0.0109 & 0.0116 & 0.0121 \\
1994 & 4.00 & (0.58,0.16,-0.53) & (0.11,-0.79,-0.08) & 0.0062 & 0.1938 & 0.0085 & 0.0121 & 0.0118 & 0.0108 \\
1981 & 4.00 & (0.26,-0.55,-0.52) & (-0.35,-0.72,-0.02) & 0.0062 & 0.1460 & 0.0081 & 0.0120 & 0.0114 & 0.0101 \\
1156 & 4.39 & (-0.16,0.21,0.38) & (0.53,-0.55,0.11) & 0.0041 & 0.1776 & 0.0079 & 0.0085 & 0.0088 & 0.0106 \\
1629 & 3.46 & (0.54,0.15,-0.45) & (-0.23,0.08,-0.73) & 0.0059 & 0.2096 & 0.0077 & 0.0089 & 0.0097 & 0.0104 \\
1923 & 4.00 & (-0.79,0.04,0.09) & (-0.75,0.28,0.02) & 0.0066 & 0.1622 & 0.0074 & 0.0188 & 0.0184 & 0.0169 \\
2011 & 4.00 & (0.79,-0.09,0.03) & (0.37,0.69,0.18) & 0.0063 & 0.2168 & 0.0064 & 0.0292 & 0.0260 & 0.0243 \\
1863 & 3.63 & (-0.45,0.30,-0.58) & (0.33,0.33,0.43) & 0.0060 & 0.1654 & 0.0063 & 0.0114 & 0.0133 & 0.0137 \\
1997 & 4.00 & (-0.76,-0.24,0.04) & (-0.00,0.14,0.79) & 0.0063 & 0.1663 & 0.0060 & 0.0114 & 0.0104 & 0.0101 \\
1983 & 4.00 & (-0.47,0.35,-0.55) & (-0.52,-0.59,0.14) & 0.0062 & 0.1721 & 0.0058 & 0.0130 & 0.0136 & 0.0144 \\
2005 & 4.00 & (0.36,0.71,0.07) & (0.48,0.64,0.06) & 0.0065 & 0.1290 & 0.0057 & 0.0136 & 0.0130 & 0.0119 \\
2081 & 4.00 & (-0.36,0.71,0.06) & (0.62,0.49,-0.14) & 0.0065 & 0.1904 & 0.0055 & 0.0104 & 0.0105 & 0.0112 \\
2048 & 4.00 & (0.80,-0.02,0.02) & (-0.26,0.41,0.64) & 0.0065 & 0.1924 & 0.0054 & 0.0092 & 0.0100 & 0.0102 \\
1579 & 3.44 & (0.21,0.46,-0.38) & (0.18,0.48,-0.59) & 0.0062 & 0.1837 & 0.0053 & 0.0137 & 0.0145 & 0.0153 \\
1986 & 4.00 & (-0.39,0.45,-0.53) & (0.11,0.03,0.79) & 0.0063 & 0.1412 & 0.0052 & 0.0123 & 0.0120 & 0.0108 \\
2007 & 4.00 & (0.77,0.22,0.04) & (0.00,0.15,-0.79) & 0.0063 & 0.1887 & 0.0051 & 0.0123 & 0.0124 & 0.0136 \\
1979 & 4.00 & (-0.53,0.00,0.60) & (-0.03,-0.12,-0.79) & 0.0067 & 0.1697 & 0.0050 & 0.0133 & 0.0132 & 0.0137 \\
2043 & 4.00 & (-0.70,-0.39,0.06) & (-0.50,0.34,0.52) & 0.0063 & 0.1638 & 0.0050 & 0.0107 & 0.0104 & 0.0105 \\
1975 & 4.00 & (0.45,0.27,0.61) & (0.04,-0.13,0.79) & 0.0071 & 0.1899 & 0.0049 & 0.0181 & 0.0190 & 0.0200 \\
1917 & 4.00 & (0.38,0.71,0.01) & (-0.68,0.36,0.20) & 0.0066 & 0.1862 & 0.0046 & 0.0101 & 0.0105 & 0.0113 \\
1972 & 4.00 & (-0.50,0.20,0.59) & (0.80,0.00,-0.06) & 0.0068 & 0.2032 & 0.0044 & 0.0094 & 0.0099 & 0.0103 \\
1974 & 4.00 & (-0.34,0.43,0.59) & (-0.00,-0.00,-0.00) & 0.0068 & 0.2039 & 0.0044 & 0.0162 & 0.0177 & 0.0190 \\
0147 & 1.00 & (0.40,0.29,-0.00) & (-0.40,-0.29,-0.00) & 0.0107 & 0.1616 & 0.0043 & 0.0104 & 0.0099 & 0.0093 \\
2015 & 4.00 & (0.57,0.56,0.03) & (0.04,-0.07,0.39) & 0.0065 & 0.1731 & 0.0042 & 0.0110 & 0.0107 & 0.0101 \\
0469 & 1.00 & (-0.16,0.78,0.03) & (0.04,-0.01,0.40) & 0.0059 & 0.1842 & 0.0039 & 0.0111 & 0.0120 & 0.0132 \\
1927 & 4.00 & (0.52,-0.61,0.01) & (0.03,0.79,0.10) & 0.0063 & 0.1738 & 0.0037 & 0.0213 & 0.0189 & 0.0169 \\
2034 & 4.00 & (-0.79,-0.07,0.06) & (0.39,0.07,-0.03) & 0.0066 & 0.1860 & 0.0036 & 0.0121 & 0.0124 & 0.0133 \\
2010 & 4.00 & (0.78,0.16,0.02) & (0.23,0.75,0.16) & 0.0065 & 0.1819 & 0.0033 & 0.0112 & 0.0101 & 0.0114 \\
1973 & 4.00 & (0.29,0.47,0.58) & (-0.08,0.07,-0.79) & 0.0067 & 0.1823 & 0.0029 & 0.0132 & 0.0138 & 0.0150 \\
1614 & 2.68 & (0.20,0.03,0.71) & (-0.11,-0.07,0.03) & 0.0063 & 0.1896 & 0.0028 & 0.0118 & 0.0121 & 0.0134 \\
1713 & 3.97 & (0.05,-0.47,0.30) & (0.68,-0.16,-0.34) & 0.0066 & 0.1842 & 0.0028 & 0.0092 & 0.0092 & 0.0100 \\
1741 & 2.77 & (0.58,-0.51,-0.06) & (-0.01,-0.05,-0.45) & 0.0061 & 0.1964 & 0.0027 & 0.0092 & 0.0098 & 0.0105 \\
2079 & 4.00 & (-0.39,-0.69,0.04) & (-0.31,0.73,-0.12) & 0.0066 & 0.1948 & 0.0027 & 0.0158 & 0.0131 & 0.0132 \\
1209 & 2.00 & (0.06,-0.01,0.85) & (-0.19,0.83,0.01) & 0.0062 & 0.1864 & 0.0026 & 0.0093 & 0.0097 & 0.0107 \\
2004 & 4.00 & (-0.27,-0.75,0.02) & (-0.22,0.77,-0.09) & 0.0063 & 0.2052 & 0.0026 & 0.0111 & 0.0107 & 0.0094 \\
2064 & 4.00 & (-0.44,-0.67,0.00) & (0.24,-0.61,-0.46) & 0.0063 & 0.1698 & 0.0023 & 0.0102 & 0.0094 & 0.0089 \\
0705 & 2.00 & (-0.03,-0.04,0.80) & (0.76,-0.26,0.02) & 0.0062 & 0.1925 & 0.0022 & 0.0105 & 0.0113 & 0.0125 \\
1095 & 2.00 & (0.22,0.77,0.02) & (-0.09,0.04,-0.79) & 0.0057 & 0.1743 & 0.0022 & 0.0124 & 0.0119 & 0.0100 \\
1659 & 3.47 & (-0.07,0.58,0.54) & (-0.04,0.17,0.43) & 0.0066 & 0.1842 & 0.0022 & 0.0111 & 0.0110 & 0.0113 \\
2058 & 4.00 & (-0.18,0.78,0.03) & (0.35,-0.33,-0.64) & 0.0062 & 0.1803 & 0.0022 & 0.0103 & 0.0097 & 0.0086 \\
1591 & 3.59 & (0.31,-0.28,0.50) & (0.48,-0.11,0.32) & 0.0066 & 0.1906 & 0.0020 & 0.0123 & 0.0121 & 0.0130 \\
1399 & 1.58 & (-0.29,-0.20,-0.23) & (-0.37,0.03,0.20) & 0.0028 & 0.2196 & 0.0019 & 0.0108 & 0.0118 & 0.0125 \\
0708 & 2.00 & (0.76,-0.23,0.04) & (-0.06,-0.10,0.79) & 0.0061 & 0.1620 & 0.0018 & 0.0108 & 0.0105 & 0.0094 \\
0968 & 2.00 & (0.07,0.80,-0.01) & (-0.60,0.51,0.10) & 0.0059 & 0.1737 & 0.0018 & 0.0088 & 0.0093 & 0.0102 \\
0888 & 2.00 & (-0.61,-0.51,0.03) & (-0.20,-0.42,0.65) & 0.0061 & 0.1805 & 0.0017 & 0.0127 & 0.0123 & 0.0102 \\
1839 & 3.76 & (0.25,-0.33,0.50) & (0.18,-0.54,-0.37) & 0.0066 & 0.1715 & 0.0017 & 0.0106 & 0.0104 & 0.0112 \\
0835 & 2.00 & (-0.48,-0.64,0.02) & (0.00,-0.00,0.00) & 0.0059 & 0.1907 & 0.0016 & 0.0097 & 0.0108 & 0.0118 \\
0900 & 2.00 & (-0.15,0.79,0.04) & (-0.30,0.38,0.64) & 0.0061 & 0.1952 & 0.0016 & 0.0086 & 0.0096 & 0.0103 \\
1532 & 3.02 & (-0.59,-0.29,0.39) & (0.17,0.12,-0.31) & 0.0065 & 0.1981 & 0.0015 & 0.0116 & 0.0114 & 0.0106 \\
1668 & 3.43 & (0.38,0.13,-0.66) & (-0.43,-0.63,0.07) & 0.0061 & 0.1434 & 0.0015 & 0.0103 & 0.0094 & 0.0082 \\
1929 & 4.00 & (0.43,-0.67,0.08) & (0.65,-0.45,0.15) & 0.0067 & 0.1835 & 0.0014 & 0.0122 & 0.0114 & 0.0097 \\
0733 & 2.00 & (0.35,-0.19,-0.02) & (-0.11,0.79,0.06) & 0.0060 & 0.2009 & 0.0012 & 0.0108 & 0.0105 & 0.0093 \\
1656 & 3.40 & (-0.37,0.19,0.59) & (-0.06,-0.08,-0.12) & 0.0066 & 0.1810 & 0.0012 & 0.0118 & 0.0124 & 0.0134 \\
0664 & 1.33 & (-0.79,-0.12,0.03) & (-0.79,-0.10,0.03) & 0.0056 & 0.1881 & 0.0011 & 0.0094 & 0.0100 & 0.0109 \\
1006 & 1.03 & (0.64,0.21,-0.35) & (-0.48,0.18,0.50) & 0.0059 & 0.1814 & 0.0011 & 0.0112 & 0.0109 & 0.0099 \\
1557 & 2.94 & (0.69,-0.07,0.20) & (-0.04,0.79,-0.13) & 0.0062 & 0.1887 & 0.0011 & 0.0137 & 0.0135 & 0.0136 \\
1696 & 2.63 & (0.67,0.33,0.09) & (-0.12,0.16,0.27) & 0.0062 & 0.1709 & 0.0011 & 0.0113 & 0.0108 & 0.0091 \\
1770 & 2.55 & (0.42,0.28,0.58) & (-0.33,0.63,-0.34) & 0.0063 & 0.1912 & 0.0011 & 0.0112 & 0.0116 & 0.0127 \\
1787 & 3.23 & (0.59,0.37,0.27) & (0.14,0.30,-0.67) & 0.0063 & 0.1954 & 0.0011 & 0.0122 & 0.0119 & 0.0108 \\
0834 & 1.00 & (-0.56,-0.57,0.03) & (-0.00,0.00,-0.00) & 0.0057 & 0.2074 & 0.0010 & 0.0099 & 0.0103 & 0.0114 \\
0907 & 1.00 & (-0.73,-0.33,-0.02) & (0.53,-0.05,0.60) & 0.0059 & 0.2097 & 0.0010 & 0.0117 & 0.0125 & 0.0135 \\
1206 & 1.00 & (0.62,-0.58,-0.05) & (0.18,0.83,0.08) & 0.0057 & 0.1797 & 0.0010 & 0.0099 & 0.0102 & 0.0112 \\
0905 & 1.00 & (0.65,-0.46,0.02) & (0.49,-0.11,0.62) & 0.0059 & 0.1925 & 0.0009 & 0.0110 & 0.0108 & 0.0116 \\
0916 & 1.00 & (-0.77,-0.21,0.01) & (-0.56,-0.57,0.08) & 0.0057 & 0.1979 & 0.0009 & 0.0097 & 0.0101 & 0.0110 \\
0966 & 2.00 & (-0.71,-0.37,0.06) & (-0.68,0.42,-0.06) & 0.0060 & 0.2054 & 0.0009 & 0.0109 & 0.0115 & 0.0125 \\
1149 & 3.00 & (0.00,-0.00,0.70) & (-0.00,-0.00,0.60) & 0.0063 & 0.1798 & 0.0009 & 0.0132 & 0.0130 & 0.0142 \\
1523 & 2.93 & (0.49,-0.26,0.46) & (0.41,0.30,0.40) & 0.0065 & 0.1963 & 0.0009 & 0.0100 & 0.0107 & 0.0117 \\
0750 & 2.00 & (-0.28,-0.48,0.57) & (0.07,-0.05,-0.80) & 0.0060 & 0.1768 & 0.0008 & 0.0109 & 0.0111 & 0.0118 \\
1000 & 1.21 & (0.31,0.63,0.34) & (-0.60,-0.02,0.48) & 0.0060 & 0.1853 & 0.0008 & 0.0105 & 0.0103 & 0.0093 \\
1086 & 1.07 & (-0.33,-0.35,0.63) & (0.59,0.18,0.16) & 0.0060 & 0.1990 & 0.0008 & 0.0112 & 0.0115 & 0.0126 \\
1197 & 2.00 & (-0.78,-0.34,-0.04) & (0.65,-0.54,0.10) & 0.0059 & 0.1903 & 0.0008 & 0.0098 & 0.0102 & 0.0111 \\
1199 & 2.00 & (0.68,-0.51,0.04) & (0.10,0.08,-0.84) & 0.0057 & 0.2148 & 0.0008 & 0.0093 & 0.0092 & 0.0101 \\
1849 & 2.70 & (0.54,-0.00,0.53) & (-0.41,0.31,0.34) & 0.0063 & 0.1963 & 0.0008 & 0.0106 & 0.0103 & 0.0094 \\
2131 & 2.00 & (0.00,0.00,0.85) & (0.00,-0.00,0.85) & 0.0060 & 0.1824 & 0.0008 & 0.0133 & 0.0142 & 0.0156 \\
0601 & 1.06 & (-0.50,0.07,0.59) & (0.02,0.04,0.66) & 0.0061 & 0.1974 & 0.0007 & 0.0091 & 0.0095 & 0.0105 \\
0635 & 1.00 & (0.67,-0.44,0.03) & (-0.06,-0.04,0.80) & 0.0059 & 0.1930 & 0.0007 & 0.0106 & 0.0111 & 0.0119 \\
0170 & 1.00 & (-0.00,-0.00,0.44) & (0.00,0.00,0.44) & 0.0071 & 0.1582 & 0.0006 & 0.0110 & 0.0109 & 0.0121 \\
0323 & 1.22 & (0.00,-0.00,0.33) & (-0.00,-0.00,-0.44) & 0.0066 & 0.1553 & 0.0006 & 0.0097 & 0.0095 & 0.0105 \\
0781 & 2.00 & (0.79,-0.14,0.03) & (0.05,0.10,-0.79) & 0.0057 & 0.2059 & 0.0006 & 0.0118 & 0.0116 & 0.0108 \\
1071 & 1.07 & (-0.13,0.20,0.66) & (0.33,-0.56,0.39) & 0.0060 & 0.1995 & 0.0006 & 0.0124 & 0.0133 & 0.0145 \\
1716 & 2.24 & (-0.29,0.36,0.53) & (-0.55,-0.08,0.46) & 0.0062 & 0.1936 & 0.0006 & 0.0095 & 0.0102 & 0.0112 \\
0256 & 2.00 & (-0.00,0.00,0.60) & (-0.00,0.00,0.60) & 0.0057 & 0.1609 & 0.0005 & 0.0116 & 0.0121 & 0.0133 \\
0351 & 1.00 & (-0.20,0.77,0.03) & (0.08,-0.01,0.80) & 0.0060 & 0.1906 & 0.0005 & 0.0115 & 0.0113 & 0.0108 \\
0936 & 2.00 & (-0.68,-0.42,-0.01) & (0.79,0.08,0.04) & 0.0060 & 0.1985 & 0.0005 & 0.0104 & 0.0110 & 0.0120 \\
0948 & 2.00 & (0.03,0.01,0.80) & (-0.42,0.38,-0.56) & 0.0061 & 0.1913 & 0.0005 & 0.0095 & 0.0094 & 0.0101 \\
1014 & 1.69 & (-0.64,0.10,0.33) & (-0.54,0.09,0.46) & 0.0061 & 0.2025 & 0.0005 & 0.0123 & 0.0121 & 0.0128 \\
1632 & 3.01 & (0.55,-0.51,0.25) & (-0.63,-0.13,-0.33) & 0.0062 & 0.1970 & 0.0005 & 0.0133 & 0.0136 & 0.0147 \\
1718 & 2.31 & (-0.30,0.33,0.61) & (-0.38,0.64,-0.09) & 0.0062 & 0.1832 & 0.0005 & 0.0108 & 0.0106 & 0.0105 \\
2006 & 4.00 & (-0.49,-0.63,-0.02) & (0.75,-0.23,0.14) & 0.0063 & 0.1810 & 0.0005 & 0.0111 & 0.0093 & 0.0085 \\
0065 & 8.00 & (-0.00,-0.00,0.50) & (0.00,0.00,0.00) & 0.0067 & 0.1400 & 0.0004 & 0.0115 & 0.0111 & 0.0094 \\
0324 & 1.22 & (-0.00,-0.00,0.33) & (-0.00,-0.00,-0.44) & 0.0088 & 0.1277 & 0.0004 & 0.0101 & 0.0098 & 0.0098 \\
0374 & 2.00 & (-0.26,0.47,0.59) & (0.00,-0.00,0.00) & 0.0062 & 0.1868 & 0.0004 & 0.0099 & 0.0103 & 0.0113 \\
0383 & 1.75 & (-0.31,0.74,0.04) & (0.10,0.01,0.79) & 0.0060 & 0.1942 & 0.0004 & 0.0108 & 0.0107 & 0.0113 \\
0476 & 1.00 & (-0.17,0.37,0.44) & (0.04,0.00,0.80) & 0.0061 & 0.1891 & 0.0004 & 0.0142 & 0.0148 & 0.0162 \\
0662 & 1.33 & (-0.68,-0.41,0.03) & (-0.02,0.09,0.79) & 0.0060 & 0.1899 & 0.0004 & 0.0099 & 0.0103 & 0.0112 \\
0688 & 1.67 & (-0.65,-0.46,0.03) & (-0.03,0.10,0.79) & 0.0061 & 0.1686 & 0.0004 & 0.0101 & 0.0098 & 0.0090 \\
0772 & 2.00 & (-0.45,0.66,0.08) & (-0.14,0.79,0.01) & 0.0060 & 0.1960 & 0.0004 & 0.0130 & 0.0137 & 0.0148 \\
0845 & 2.00 & (0.74,-0.30,0.04) & (-0.04,-0.05,0.40) & 0.0061 & 0.2014 & 0.0004 & 0.0130 & 0.0138 & 0.0148 \\
0941 & 2.00 & (-0.03,0.03,0.80) & (-0.05,-0.57,-0.56) & 0.0062 & 0.1785 & 0.0004 & 0.0094 & 0.0100 & 0.0111 \\
0988 & 2.00 & (-0.04,0.03,0.80) & (-0.20,-0.77,0.02) & 0.0062 & 0.1619 & 0.0004 & 0.0090 & 0.0096 & 0.0106 \\
1090 & 1.59 & (-0.30,-0.33,0.48) & (-0.30,-0.34,0.52) & 0.0061 & 0.1891 & 0.0004 & 0.0117 & 0.0126 & 0.0139 \\
1529 & 3.14 & (0.33,-0.59,0.13) & (-0.38,0.50,0.48) & 0.0063 & 0.1636 & 0.0004 & 0.0102 & 0.0095 & 0.0079 \\
1642 & 3.29 & (-0.30,-0.54,0.26) & (0.72,-0.15,0.04) & 0.0066 & 0.1904 & 0.0004 & 0.0137 & 0.0132 & 0.0112 \\
1676 & 3.25 & (0.11,0.18,0.44) & (0.31,-0.13,0.22) & 0.0066 & 0.2032 & 0.0004 & 0.0108 & 0.0106 & 0.0107 \\
1692 & 2.88 & (-0.51,0.31,-0.02) & (-0.34,-0.41,-0.07) & 0.0060 & 0.1886 & 0.0004 & 0.0111 & 0.0108 & 0.0094 \\
0333 & 2.00 & (0.00,0.00,0.80) & (-0.00,0.00,0.80) & 0.0063 & 0.1786 & 0.0003 & 0.0120 & 0.0129 & 0.0143 \\
0348 & 1.19 & (-0.21,0.45,0.60) & (0.06,0.01,0.76) & 0.0061 & 0.1781 & 0.0003 & 0.0116 & 0.0114 & 0.0120 \\
0478 & 1.32 & (-0.23,0.63,0.14) & (0.08,-0.00,0.78) & 0.0060 & 0.1859 & 0.0003 & 0.0134 & 0.0131 & 0.0118 \\
0571 & 1.09 & (0.00,0.08,-0.02) & (-0.00,0.00,-0.29) & 0.0057 & 0.1986 & 0.0003 & 0.0091 & 0.0094 & 0.0104 \\
0575 & 1.20 & (-0.00,0.01,0.39) & (0.00,-0.00,0.14) & 0.0059 & 0.1901 & 0.0003 & 0.0105 & 0.0111 & 0.0123 \\
0691 & 1.67 & (-0.73,-0.31,0.06) & (0.02,0.80,-0.07) & 0.0059 & 0.1792 & 0.0003 & 0.0104 & 0.0101 & 0.0088 \\
0745 & 2.00 & (-0.16,-0.52,0.59) & (-0.00,0.00,0.00) & 0.0062 & 0.2040 & 0.0003 & 0.0087 & 0.0094 & 0.0103 \\
0830 & 2.00 & (0.70,-0.38,0.06) & (-0.65,-0.46,-0.08) & 0.0059 & 0.1802 & 0.0003 & 0.0123 & 0.0118 & 0.0099 \\
0859 & 1.00 & (-0.01,0.04,0.80) & (-0.34,-0.21,0.01) & 0.0060 & 0.1943 & 0.0003 & 0.0128 & 0.0126 & 0.0128 \\
0991 & 2.00 & (-0.67,-0.44,0.03) & (-0.32,-0.72,0.12) & 0.0060 & 0.2002 & 0.0003 & 0.0121 & 0.0131 & 0.0142 \\
1011 & 1.53 & (0.51,0.31,0.31) & (0.33,0.48,0.48) & 0.0060 & 0.1857 & 0.0003 & 0.0126 & 0.0132 & 0.0145 \\
1020 & 1.24 & (0.37,0.38,0.49) & (0.53,-0.27,0.52) & 0.0060 & 0.1910 & 0.0003 & 0.0095 & 0.0100 & 0.0108 \\
1023 & 1.22 & (-0.59,-0.02,0.36) & (0.21,-0.63,-0.37) & 0.0057 & 0.1880 & 0.0003 & 0.0089 & 0.0093 & 0.0102 \\
1063 & 1.78 & (-0.47,0.28,-0.29) & (-0.29,-0.41,0.58) & 0.0059 & 0.1965 & 0.0003 & 0.0097 & 0.0103 & 0.0113 \\
1070 & 1.20 & (-0.15,-0.43,0.64) & (-0.42,-0.41,-0.49) & 0.0059 & 0.1691 & 0.0003 & 0.0107 & 0.0111 & 0.0122 \\
1571 & 3.44 & (-0.28,-0.35,0.64) & (0.48,-0.51,-0.10) & 0.0068 & 0.1777 & 0.0003 & 0.0087 & 0.0093 & 0.0102 \\
1616 & 2.87 & (0.53,0.30,0.42) & (-0.38,0.02,-0.54) & 0.0062 & 0.2089 & 0.0003 & 0.0128 & 0.0137 & 0.0149 \\
1709 & 3.44 & (-0.09,0.20,0.29) & (0.14,0.47,-0.60) & 0.0065 & 0.1871 & 0.0003 & 0.0101 & 0.0098 & 0.0097 \\
1930 & 4.00 & (0.09,0.79,0.04) & (-0.17,0.07,-0.78) & 0.0065 & 0.2064 & 0.0003 & 0.0103 & 0.0089 & 0.0081 \\
2161 & 3.00 & (0.00,-0.00,0.60) & (0.00,0.00,0.00) & 0.0057 & 0.1913 & 0.0003 & 0.0118 & 0.0116 & 0.0125 \\
0178 & 1.00 & (0.00,0.00,0.99) & (-0.00,-0.00,0.99) & 0.0056 & 0.1941 & 0.0002 & 0.0144 & 0.0151 & 0.0165 \\
0372 & 1.50 & (0.00,-0.00,0.80) & (-0.00,0.00,-0.40) & 0.0060 & 0.1851 & 0.0002 & 0.0088 & 0.0093 & 0.0102 \\
0395 & 1.00 & (-0.10,0.42,-0.42) & (0.04,-0.01,0.80) & 0.0059 & 0.2164 & 0.0002 & 0.0095 & 0.0101 & 0.0110 \\
0408 & 2.00 & (-0.29,0.53,0.02) & (0.08,0.01,0.80) & 0.0061 & 0.1909 & 0.0002 & 0.0113 & 0.0123 & 0.0135 \\
0505 & 1.85 & (-0.26,0.51,0.09) & (0.08,0.01,0.79) & 0.0061 & 0.2000 & 0.0002 & 0.0090 & 0.0095 & 0.0104 \\
0655 & 1.33 & (0.61,-0.52,0.03) & (0.00,0.00,-0.00) & 0.0057 & 0.2062 & 0.0002 & 0.0113 & 0.0119 & 0.0130 \\
0679 & 1.67 & (-0.04,-0.04,0.80) & (0.79,-0.14,0.03) & 0.0062 & 0.1587 & 0.0002 & 0.0118 & 0.0124 & 0.0137 \\
0774 & 2.00 & (0.79,-0.12,0.02) & (-0.00,-0.00,0.00) & 0.0059 & 0.1851 & 0.0002 & 0.0107 & 0.0104 & 0.0089 \\
0777 & 2.00 & (0.80,0.05,0.07) & (0.75,-0.29,0.01) & 0.0059 & 0.2001 & 0.0002 & 0.0127 & 0.0125 & 0.0120 \\
0870 & 1.00 & (0.04,-0.01,0.80) & (-0.02,0.40,0.01) & 0.0060 & 0.1930 & 0.0002 & 0.0103 & 0.0106 & 0.0117 \\
0957 & 2.00 & (-0.73,-0.33,0.00) & (0.58,-0.07,-0.54) & 0.0059 & 0.1765 & 0.0002 & 0.0110 & 0.0105 & 0.0090 \\
0963 & 1.00 & (0.58,-0.55,-0.00) & (-0.57,0.56,0.00) & 0.0057 & 0.2067 & 0.0002 & 0.0108 & 0.0105 & 0.0097 \\
1084 & 1.76 & (0.48,-0.16,0.48) & (-0.73,0.10,0.30) & 0.0061 & 0.1841 & 0.0002 & 0.0131 & 0.0128 & 0.0116 \\
1196 & 1.00 & (0.64,-0.55,0.06) & (0.64,-0.55,0.06) & 0.0057 & 0.1681 & 0.0002 & 0.0110 & 0.0115 & 0.0126 \\
1406 & 1.60 & (-0.29,0.29,0.24) & (-0.38,-0.01,0.15) & 0.0028 & 0.1984 & 0.0002 & 0.0103 & 0.0110 & 0.0120 \\
1495 & 1.00 & (-0.00,-0.00,0.78) & (0.00,0.00,0.53) & 0.0061 & 0.1920 & 0.0002 & 0.0124 & 0.0132 & 0.0144 \\
1518 & 2.08 & (0.23,-0.66,0.08) & (-0.60,-0.17,0.15) & 0.0059 & 0.1989 & 0.0002 & 0.0109 & 0.0113 & 0.0123 \\
1645 & 2.29 & (-0.33,-0.38,0.45) & (0.00,-0.64,0.46) & 0.0062 & 0.1917 & 0.0002 & 0.0115 & 0.0113 & 0.0113 \\
1852 & 3.03 & (-0.45,0.25,0.46) & (-0.15,0.71,-0.25) & 0.0063 & 0.1921 & 0.0002 & 0.0096 & 0.0099 & 0.0108 \\
1860 & 3.42 & (-0.35,-0.20,0.64) & (0.67,0.33,-0.24) & 0.0067 & 0.1779 & 0.0002 & 0.0106 & 0.0112 & 0.0123 \\
2097 & 1.00 & (-0.00,0.00,0.30) & (0.00,-0.00,-0.00) & 0.0051 & 0.2019 & 0.0002 & 0.0098 & 0.0105 & 0.0116 \\
2125 & 2.00 & (0.00,-0.00,0.30) & (0.00,-0.00,0.30) & 0.0056 & 0.2023 & 0.0002 & 0.0105 & 0.0104 & 0.0115 \\
0155 & 1.00 & (-0.00,-0.00,0.80) & (0.00,0.00,0.80) & 0.0055 & 0.1886 & 0.0001 & 0.0123 & 0.0130 & 0.0143 \\
0179 & 1.50 & (-0.00,-0.00,0.99) & (0.13,0.05,0.14) & 0.0056 & 0.1677 & 0.0001 & 0.0126 & 0.0130 & 0.0144 \\
0328 & 1.00 & (0.00,0.00,0.80) & (-0.00,-0.00,0.80) & 0.0062 & 0.1871 & 0.0001 & 0.0100 & 0.0105 & 0.0115 \\
0415 & 1.00 & (0.00,0.00,-0.00) & (-0.00,0.00,-0.40) & 0.0056 & 0.1978 & 0.0001 & 0.0089 & 0.0092 & 0.0102 \\
0495 & 1.38 & (-0.03,0.13,0.01) & (0.01,-0.00,0.40) & 0.0059 & 0.1862 & 0.0001 & 0.0094 & 0.0100 & 0.0111 \\
0564 & 1.69 & (-0.01,0.01,0.27) & (0.00,0.00,0.61) & 0.0061 & 0.1912 & 0.0001 & 0.0089 & 0.0091 & 0.0101 \\
0681 & 1.67 & (0.06,-0.01,0.80) & (-0.28,0.75,0.03) & 0.0062 & 0.1897 & 0.0001 & 0.0122 & 0.0129 & 0.0142 \\
0694 & 1.67 & (-0.07,0.80,0.03) & (0.10,-0.02,0.79) & 0.0060 & 0.1938 & 0.0001 & 0.0119 & 0.0118 & 0.0124 \\
0706 & 2.00 & (-0.02,0.05,0.80) & (-0.40,-0.69,0.03) & 0.0063 & 0.1693 & 0.0001 & 0.0111 & 0.0116 & 0.0128 \\
0752 & 2.00 & (-0.39,0.42,0.56) & (-0.48,-0.63,0.12) & 0.0062 & 0.1830 & 0.0001 & 0.0107 & 0.0105 & 0.0109 \\
0854 & 2.00 & (0.70,-0.39,0.04) & (0.36,-0.18,0.02) & 0.0059 & 0.1927 & 0.0001 & 0.0138 & 0.0135 & 0.0121 \\
0915 & 2.00 & (0.71,-0.36,0.08) & (-0.30,-0.74,-0.07) & 0.0059 & 0.1865 & 0.0001 & 0.0108 & 0.0104 & 0.0091 \\
0964 & 2.00 & (0.70,-0.38,0.01) & (-0.73,0.33,-0.01) & 0.0060 & 0.1770 & 0.0001 & 0.0116 & 0.0112 & 0.0093 \\
1044 & 1.77 & (0.66,0.08,-0.25) & (-0.03,-0.74,0.26) & 0.0057 & 0.2069 & 0.0001 & 0.0090 & 0.0095 & 0.0104 \\
1068 & 1.46 & (0.08,0.02,0.19) & (-0.63,0.02,0.43) & 0.0060 & 0.1909 & 0.0001 & 0.0095 & 0.0100 & 0.0111 \\
1194 & 2.00 & (0.75,-0.39,0.07) & (-0.68,-0.49,-0.09) & 0.0059 & 0.1840 & 0.0001 & 0.0104 & 0.0100 & 0.0083 \\
1477 & 1.00 & (-0.00,-0.00,0.80) & (0.00,0.00,0.80) & 0.0062 & 0.1879 & 0.0001 & 0.0094 & 0.0100 & 0.0110 \\
1521 & 3.07 & (-0.38,0.26,0.39) & (0.36,0.44,0.28) & 0.0065 & 0.1469 & 0.0001 & 0.0097 & 0.0103 & 0.0114 \\
1747 & 2.66 & (-0.21,-0.00,0.70) & (-0.09,0.16,-0.39) & 0.0065 & 0.1829 & 0.0001 & 0.0119 & 0.0122 & 0.0135 \\
1893 & 2.62 & (0.30,0.51,0.51) & (-0.31,-0.17,0.71) & 0.0065 & 0.1882 & 0.0001 & 0.0136 & 0.0134 & 0.0128 \\
2156 & 3.00 & (-0.00,0.00,0.40) & (-0.00,0.00,-0.60) & 0.0059 & 0.1788 & 0.0001 & 0.0103 & 0.0101 & 0.0106 \\
0255 & 2.00 & (0.00,-0.00,0.60) & (0.00,-0.00,-0.00) & 0.0056 & 0.1788 & 0.0000 & 0.0110 & 0.0114 & 0.0125 \\
0418 & 1.00 & (0.00,0.00,-0.00) & (-0.00,-0.00,0.40) & 0.0059 & 0.1858 & 0.0000 & 0.0090 & 0.0094 & 0.0105 \\
0553 & 1.07 & (-0.01,0.03,0.69) & (0.00,0.00,0.46) & 0.0061 & 0.1864 & 0.0000 & 0.0091 & 0.0096 & 0.0106 \\
0581 & 1.68 & (-0.21,0.55,0.50) & (0.01,0.00,0.07) & 0.0061 & 0.1932 & 0.0000 & 0.0117 & 0.0119 & 0.0132 \\
0607 & 1.50 & (-0.04,0.20,0.22) & (-0.12,0.37,0.22) & 0.0060 & 0.1865 & 0.0000 & 0.0098 & 0.0103 & 0.0112 \\
2101 & 1.00 & (-0.00,0.00,0.60) & (0.00,-0.00,0.00) & 0.0052 & 0.2245 & 0.0000 & 0.0104 & 0.0111 & 0.0123 \\

%% file: draft.bbl
\begin{thebibliography}{10}

\bibitem{2021arXiv211103606T}
{The LIGO Scientific Collaboration}, {the Virgo Collaboration}, {the KAGRA Collaboration}, R.~{Abbott}, et~al.
\newblock {GWTC-3: Compact Binary Coalescences Observed by LIGO and Virgo During the Second Part of the Third Observing Run}.
\newblock {\em arXiv e-prints}, page arXiv:2111.03606, November 2021.

\bibitem{2023arXiv230203676T}
{The LIGO Scientific Collaboration}, {the Virgo Collaboration}, {the KAGRA Collaboration}, R.~{Abbott}, et~al.
\newblock {Open data from the third observing run of LIGO, Virgo, KAGRA and GEO}.
\newblock {\em arXiv e-prints}, page arXiv:2302.03676, February 2023.

\bibitem{Miller2019}
M.~Coleman Miller and Nicols Yunes.
\newblock The new frontier of gravitational waves.
\newblock {\em Nature}, 568(7753):469--476, Apr 2019.

\bibitem{2023LRR....26....2A}
Pau {Amaro-Seoane} et~al.
\newblock {Astrophysics with the Laser Interferometer Space Antenna}.
\newblock {\em Living Reviews in Relativity}, 26(1):2, December 2023.

\bibitem{2007arXiv0711.1115J}
Piotr {Jaranowski} and Andrzej {Kr{\'o}lak}.
\newblock {Gravitational-Wave Data Analysis. Formalism and Sample Applications: The Gaussian Case}.
\newblock {\em arXiv e-prints}, page arXiv:0711.1115, November 2007.

\bibitem{2021hgwa.bookE..43K}
Andrzej {Kr{\'o}lak}.
\newblock {Principles of Gravitational-Wave Data Analysis}.
\newblock In {\em Handbook of Gravitational Wave Astronomy}, page~43. 2021.

\bibitem{2022NatAs...6.1356S}
Lorenzo {Speri}, Nikolaos {Karnesis}, Arianna~I. {Renzini}, and Jonathan~R. {Gair}.
\newblock {A roadmap of gravitational wave data analysis}.
\newblock {\em Nature Astronomy}, 6:1356--1363, December 2022.

\bibitem{RevModPhys.94.025001}
Nelson Christensen and Renate Meyer.
\newblock Parameter estimation with gravitational waves.
\newblock {\em Rev. Mod. Phys.}, 94:025001, Apr 2022.

\bibitem{2017arXiv171009256C}
Eric {Chassande-Mottin}, Eric {Lebigot}, Hugo {Magaldi}, Eve {Chase}, Archana {Pai}, Gayathri {V}, and Gabriele {Vedovato}.
\newblock {Wavelet graphs for the direct detection of gravitational waves}.
\newblock {\em arXiv e-prints}, page arXiv:1710.09256, October 2017.

\bibitem{PhysRevD.98.024028}
P.~Bacon, V.~Gayathri, E.~Chassande-Mottin, A.~Pai, F.~Salemi, and G.~Vedovato.
\newblock Driving unmodeled gravitational-wave transient searches using astrophysical information.
\newblock {\em Phys. Rev. D}, 98:024028, Jul 2018.

\bibitem{2020arXiv200503745C}
Elena {Cuoco}, Jade {Powell}, Marco {Cavagli{\`a}}, Kendall {Ackley}, Michal {Bejger}, Chayan {Chatterjee}, Michael {Coughlin}, Scott {Coughlin}, Paul {Easter}, Reed {Essick}, Hunter {Gabbard}, Timothy {Gebhard}, Shaon {Ghosh}, Leila {Haegel}, Alberto {Iess}, David {Keitel}, Zsuzsa {Marka}, Szabolcs {Marka}, Filip {Morawski}, Tri {Nguyen}, Rich {Ormiston}, Michael {Puerrer}, Massimiliano {Razzano}, Kai {Staats}, Gabriele {Vajente}, and Daniel {Williams}.
\newblock {Enhancing Gravitational-Wave Science with Machine Learning}.
\newblock {\em arXiv e-prints}, page arXiv:2005.03745, May 2020.

\bibitem{2023PhRvD.107b3021S}
Marlin~B. {Sch{\"a}fer}, Ond{\v{r}}ej {Zelenka}, Alexander~H. {Nitz}, He~{Wang}, Shichao {Wu}, Zong-Kuan {Guo}, Zhoujian {Cao}, Zhixiang {Ren}, Paraskevi {Nousi}, Nikolaos {Stergioulas}, Panagiotis {Iosif}, Alexandra~E. {Koloniari}, Anastasios {Tefas}, Nikolaos {Passalis}, Francesco {Salemi}, Gabriele {Vedovato}, Sergey {Klimenko}, Tanmaya {Mishra}, Bernd {Br{\"u}gmann}, Elena {Cuoco}, E.~A. {Huerta}, Chris {Messenger}, and Frank {Ohme}.
\newblock {First machine learning gravitational-wave search mock data challenge}.
\newblock {\em \prd}, 107(2):023021, January 2023.

\bibitem{PhysRevD.101.104003}
He~Wang, Shichao Wu, Zhoujian Cao, Xiaolin Liu, and Jian-Yang Zhu.
\newblock Gravitational-wave signal recognition of ligo data by deep learning.
\newblock {\em Phys. Rev. D}, 101:104003, May 2020.

\bibitem{PhysRevD.103.024040}
Heming Xia, Lijing Shao, Junjie Zhao, and Zhoujian Cao.
\newblock Improved deep learning techniques in gravitational-wave data analysis.
\newblock {\em Phys. Rev. D}, 103:024040, Jan 2021.

\bibitem{PhysRevD.105.083013}
CunLiang Ma, Wei Wang, He~Wang, and Zhoujian Cao.
\newblock Ensemble of deep convolutional neural networks for real-time gravitational wave signal recognition.
\newblock {\em Phys. Rev. D}, 105:083013, Apr 2022.

\bibitem{PhysRevD.107.063029}
Cunliang Ma, Wei Wang, He~Wang, and Zhoujian Cao.
\newblock Artificial intelligence model for gravitational wave search based on the waveform envelope.
\newblock {\em Phys. Rev. D}, 107:063029, Mar 2023.

\bibitem{PhysRevD.101.084002}
B.~P. Abbott et~al.
\newblock Optically targeted search for gravitational waves emitted by core-collapse supernovae during the first and second observing runs of advanced ligo and advanced virgo.
\newblock {\em Phys. Rev. D}, 101:084002, Apr 2020.

\bibitem{PhysRevLett.95.121101}
Frans Pretorius.
\newblock Evolution of binary black-hole spacetimes.
\newblock {\em Phys. Rev. Lett.}, 95:121101, Sep 2005.

\bibitem{PhysRevLett.96.111101}
M.~Campanelli, C.~O. Lousto, P.~Marronetti, and Y.~Zlochower.
\newblock Accurate evolutions of orbiting black-hole binaries without excision.
\newblock {\em Phys. Rev. Lett.}, 96:111101, Mar 2006.

\bibitem{PhysRevLett.96.111102}
John~G. Baker, Joan Centrella, Dae-Il Choi, Michael Koppitz, and James van Meter.
\newblock Gravitational-wave extraction from an inspiraling configuration of merging black holes.
\newblock {\em Phys. Rev. Lett.}, 96:111102, Mar 2006.

\bibitem{PhysRevD.78.124011}
Zhoujian Cao, Hwei-Jang Yo, and Jui-Ping Yu.
\newblock Reinvestigation of moving punctured black holes with a new code.
\newblock {\em Phys. Rev. D}, 78:124011, Dec 2008.

\bibitem{Zhao2020}
Tianyu Zhao, Zhoujian Cao, Chun-Yu Lin, and Hwei-Jang Yo.
\newblock {\em Numerical Relativity for Gravitational Wave Source Modelling}, pages 1--30.
\newblock Springer Singapore, Singapore, 2020.

\bibitem{PhysRevD.49.2658}
Curt Cutler and \'Eanna~E. Flanagan.
\newblock Gravitational waves from merging compact binaries: How accurately can one extract the binary's parameters from the inspiral waveform?
\newblock {\em Phys. Rev. D}, 49:2658--2697, Mar 1994.

\bibitem{PhysRevD.59.084006}
A.~Buonanno and T.~Damour.
\newblock Effective one-body approach to general relativistic two-body dynamics.
\newblock {\em Phys. Rev. D}, 59:084006, Mar 1999.

\bibitem{PhysRevD.76.104049}
Alessandra Buonanno, Yi~Pan, John~G. Baker, Joan Centrella, Bernard~J. Kelly, Sean~T. McWilliams, and James~R. van Meter.
\newblock Approaching faithful templates for nonspinning binary black holes using the effective-one-body approach.
\newblock {\em Phys. Rev. D}, 76:104049, Nov 2007.

\bibitem{PhysRevD.95.044028_SEOBNRv4}
Alejandro Boh\'e, Lijing Shao, Andrea Taracchini, Alessandra Buonanno, Stanislav Babak, Ian~W. Harry, Ian Hinder, Serguei Ossokine, Michael P\"urrer, Vivien Raymond, Tony Chu, Heather Fong, Prayush Kumar, Harald~P. Pfeiffer, Michael Boyle, Daniel~A. Hemberger, Lawrence~E. Kidder, Geoffrey Lovelace, Mark~A. Scheel, and B\'ela Szil\'agyi.
\newblock Improved effective-one-body model of spinning, nonprecessing binary black holes for the era of gravitational-wave astrophysics with advanced detectors.
\newblock {\em Phys. Rev. D}, 95:044028, Feb 2017.

\bibitem{PhysRevD.101.101501_TEOBeccc}
Danilo Chiaramello and Alessandro Nagar.
\newblock Faithful analytical effective-one-body waveform model for spin-aligned, moderately eccentric, coalescing black hole binaries.
\newblock {\em Phys. Rev. D}, 101:101501, May 2020.

\bibitem{PhysRevD.96.044028}
Zhoujian Cao and Wen-Biao Han.
\newblock Waveform model for an eccentric binary black hole based on the effective-one-body-numerical-relativity formalism.
\newblock {\em Phys. Rev. D}, 96:044028, Aug 2017.

\bibitem{PhysRevD.101.044049}
Xiaolin Liu, Zhoujian Cao, and Lijing Shao.
\newblock Validating the effective-one-body numerical-relativity waveform models for spin-aligned binary black holes along eccentric orbits.
\newblock {\em Phys. Rev. D}, 101:044049, Feb 2020.

\bibitem{2022CQGra..39c5009L}
Xiaolin {Liu}, Zhoujian {Cao}, and Zong-Hong {Zhu}.
\newblock {A higher-multipole gravitational waveform model for an eccentric binary black holes based on the effective-one-body-numerical-relativity formalism}.
\newblock {\em Classical and Quantum Gravity}, 39(3):035009, February 2022.

\bibitem{IJMPD.32.2350015}
Xiaolin Liu, Zhoujian Cao, and Lijing Shao.
\newblock Upgraded waveform model of eccentric binary black hole based on effective-one-body-numerical-relativity for spin-aligned binary black holes.
\newblock {\em International Journal of Modern Physics D}, 32:2350015, Feb 2023.

\bibitem{2020PhRvD.102f4001P}
Geraint {Pratten}, Sascha {Husa}, Cecilio {Garc{\'\i}a-Quir{\'o}s}, Marta {Colleoni}, Antoni {Ramos-Buades}, H{\'e}ctor {Estell{\'e}s}, and Rafel {Jaume}.
\newblock {Setting the cornerstone for a family of models for gravitational waves from compact binaries: The dominant harmonic for nonprecessing quasicircular black holes}.
\newblock {\em \prd}, 102(6):064001, September 2020.

\bibitem{PhysRevD.102.064002}
Cecilio Garc\'{\i}a-Quir\'os, Marta Colleoni, Sascha Husa, H\'ector Estell\'es, Geraint Pratten, Antoni Ramos-Buades, Maite Mateu-Lucena, and Rafel Jaume.
\newblock Multimode frequency-domain model for the gravitational wave signal from nonprecessing black-hole binaries.
\newblock {\em Phys. Rev. D}, 102:064002, Sep 2020.

\bibitem{PhysRevD.96.024058}
Jonathan Blackman, Scott~E. Field, Mark~A. Scheel, Chad~R. Galley, Christian~D. Ott, Michael Boyle, Lawrence~E. Kidder, Harald~P. Pfeiffer, and B\'ela Szil\'agyi.
\newblock Numerical relativity waveform surrogate model for generically precessing binary black hole mergers.
\newblock {\em Phys. Rev. D}, 96:024058, Jul 2017.

\bibitem{PhysRevResearch.1.033015}
Vijay Varma, Scott~E. Field, Mark~A. Scheel, Jonathan Blackman, Davide Gerosa, Leo~C. Stein, Lawrence~E. Kidder, and Harald~P. Pfeiffer.
\newblock Surrogate models for precessing binary black hole simulations with unequal masses.
\newblock {\em Phys. Rev. Res.}, 1:033015, Oct 2019.

\bibitem{PhysRevD.103.064022}
Tousif Islam, Vijay Varma, Jackie Lodman, Scott~E. Field, Gaurav Khanna, Mark~A. Scheel, Harald~P. Pfeiffer, Davide Gerosa, and Lawrence~E. Kidder.
\newblock Eccentric binary black hole surrogate models for the gravitational waveform and remnant properties: Comparable mass, nonspinning case.
\newblock {\em Phys. Rev. D}, 103:064022, Mar 2021.

\bibitem{2019CQGra..36s5006B}
Michael {Boyle}, Daniel {Hemberger}, Dante A.~B. {Iozzo}, Geoffrey {Lovelace}, Serguei {Ossokine}, Harald~P. {Pfeiffer}, Mark~A. {Scheel}, Leo~C. {Stein}, Charles~J. {Woodford}, Aaron~B. {Zimmerman}, Nousha {Afshari}, Kevin {Barkett}, Jonathan {Blackman}, Katerina {Chatziioannou}, Tony {Chu}, Nicholas {Demos}, Nils {Deppe}, Scott~E. {Field}, Nils~L. {Fischer}, Evan {Foley}, Heather {Fong}, Alyssa {Garcia}, Matthew {Giesler}, Francois {Hebert}, Ian {Hinder}, Reza {Katebi}, Haroon {Khan}, Lawrence~E. {Kidder}, Prayush {Kumar}, Kevin {Kuper}, Halston {Lim}, Maria {Okounkova}, Teresita {Ramirez}, Samuel {Rodriguez}, Hannes~R. {R{\"u}ter}, Patricia {Schmidt}, Bela {Szilagyi}, Saul~A. {Teukolsky}, Vijay {Varma}, and Marissa {Walker}.
\newblock {The SXS collaboration catalog of binary black hole simulations}.
\newblock {\em Classical and Quantum Gravity}, 36(19):195006, October 2019.

\bibitem{PyCBC}
{LVK collaboration}.
\newblock Pycbc software.
\newblock \url{https://pycbc.org/}.

\bibitem{SXSBBH}
Caltech-Cornell-CITA.
\newblock binary black hole simulation results.
\newblock \url{http://www.black-holes.org/waveforms}.

\bibitem{chu2016accuracy}
Tony Chu, Heather Fong, Prayush Kumar, Harald~P Pfeiffer, Michael Boyle, Daniel~A Hemberger, Lawrence~E Kidder, Mark~A Scheel, and Bela Szilagyi.
\newblock On the accuracy and precision of numerical waveforms: Effect of waveform extraction methodology.
\newblock {\em Classical and Quantum Gravity}, 33(16):165001, 2016.

\bibitem{mckechan2010tapering}
DJA McKechan, C~Robinson, and Bangalore~Suryanarayana Sathyaprakash.
\newblock A tapering window for time-domain templates and simulated signals in the detection of gravitational waves from coalescing compact binaries.
\newblock {\em Classical and Quantum Gravity}, 27(8):084020, 2010.

\bibitem{Sho10}
D~Shoemaker (LIGO Scientific~Collaboration).
\newblock 2010 advanced ligo anticipated sensitivity curves ligo document t0900288-v3.
\newblock {\em URL https://dcc.ligo.org/cgi-bin/DocDB/ShowDocument?docid=2974}, 2010.

\bibitem{PhysRevLett.118.171101}
M.~Armano et~al.
\newblock Charge-induced force noise on free-falling test masses: Results from lisa pathfinder.
\newblock {\em Phys. Rev. Lett.}, 118:171101, Apr 2017.

\bibitem{Ruan2020}
Wen-Hong Ruan, Chang Liu, Zong-Kuan Guo, Yue-Liang Wu, and Rong-Gen Cai.
\newblock The lisa-taiji network.
\newblock {\em Nature Astronomy}, 4(2):108--109, Feb 2020.

\bibitem{luo2016tianqin}
Jun Luo, Li-Sheng Chen, Hui-Zong Duan, Yun-Gui Gong, Shoucun Hu, Jianghui Ji, Qi~Liu, Jianwei Mei, Vadim Milyukov, Mikhail Sazhin, et~al.
\newblock Tianqin: a space-borne gravitational wave detector.
\newblock {\em Classical and Quantum Gravity}, 33(3):035010, 2016.

\bibitem{Luo_2020}
Jun Luo, Yan-Zheng Bai, Lin Cai, Bin Cao, Wei-Ming Chen, Yu~Chen, De-Cong Cheng, Yan-Wei Ding, Hui-Zong Duan, Xingyu Gou, Chao-Zheng Gu, De-Feng Gu, Zi-Qi He, Shuang Hu, Yuexin Hu, Xiang-Qing Huang, Qinghua Jiang, Yuan-Ze Jiang, Hong-Gang Li, Hong-Yin Li, Jia Li, Ming Li, Zhu Li, Zhu-Xi Li, Yu-Rong Liang, Fang-Jie Liao, Yan-Chong Liu, Li~Liu, Pei-Bo Liu, Xuhui Liu, Yuan Liu, Xiong-Fei Lu, Yan Luo, Jianwei Mei, Min Ming, Shao-Bo Qu, Ding-Yin Tan, Mi~Tang, Liang-Cheng Tu, Cheng-Rui Wang, Fengbin Wang, Guan-Fang Wang, Jian Wang, Lijiao Wang, Xudong Wang, Ran Wei, Shu-Chao Wu, Chun-Yu Xiao, Meng-Zhe Xie, Xiao-Shi Xu, Liang Yang, Ming-Lin Yang, Shan-Qing Yang, Hsien-Chi Yeh, Jian-Bo Yu, Lihua Zhang, Meng-Hao Zhao, and Ze-Bing Zhou.
\newblock The first round result from the {TianQin}-1 satellite.
\newblock {\em Classical and Quantum Gravity}, 37(18):185013, aug 2020.

\bibitem{PhysRevD.102.124037}
Alexandre Toubiana, Sylvain Marsat, Stanislav Babak, John Baker, and Tito Dal~Canton.
\newblock Parameter estimation of stellar-mass black hole binaries with lisa.
\newblock {\em Phys. Rev. D}, 102:124037, Dec 2020.

\bibitem{Robson_2019}
Travis Robson, Neil~J Cornish, and Chang Liu.
\newblock The construction and use of {LISA} sensitivity curves.
\newblock {\em Classical and Quantum Gravity}, 36(10):105011, apr 2019.

\bibitem{2020IJMPA..3550075R}
Wen-Hong {Ruan}, Zong-Kuan {Guo}, Rong-Gen {Cai}, and Yuan-Zhong {Zhang}.
\newblock {Taiji program: Gravitational-wave sources}.
\newblock {\em International Journal of Modern Physics A}, 35(17):2050075, June 2020.

\bibitem{Cornish_2017}
Neil Cornish and Travis Robson.
\newblock Galactic binary science with the new lisa design.
\newblock {\em Journal of Physics: Conference Series}, 840(1):012024, may 2017.

\bibitem{2014CQGra..31k5004A}
J.~{Aasi} et~al.
\newblock {The NINJA-2 project: detecting and characterizing gravitational waveforms modelled using numerical binary black hole simulations}.
\newblock {\em Classical and Quantum Gravity}, 31(11):115004, June 2014.

\end{thebibliography}
